\documentclass{emulateapj}
\usepackage{apjfonts,psfig}

\newcommand{\pks}{PKS~0637$-$752}
\newcommand{\lae}{\mathrel{\raise .4ex\hbox{\rlap{$<$}\lower 1.2ex\hbox{$\sim$}}}}
\newcommand{\gae}{\mathrel{\raise .4ex\hbox{\rlap{$>$}\lower 1.2ex\hbox{$\sim$}}}}

\slugcomment{Submitted to Ap. J.}
\shorttitle{X-rays from Quasar Jets}
\shortauthors{Marshall et al.}

\begin{document}

\title{A Chandra Survey of Quasar Jets: First Results}

\author{H. L. Marshall\altaffilmark{1},
D.A. Schwartz\altaffilmark{2},
J.E.J. Lovell\altaffilmark{3},
D.W. Murphy\altaffilmark{4},
D.M. Worrall\altaffilmark{2,5},
M. Birkinshaw\altaffilmark{2,5},
J.M. Gelbord\altaffilmark{1},
E.S. Perlman\altaffilmark{6},
D.L. Jauncey\altaffilmark{3}}
\altaffiltext{1}{Center for Space Research, Massachusetts Institute of
	Technology, 77 Massachusetts Ave., Cambridge, MA 02139, USA}
\altaffiltext{2}{Harvard-Smithsonian Center for Astrophysics,
	60 Garden St., Cambridge, MA 02138, USA}
\altaffiltext{3}{CSIRO Australia Telescope National Facility,
	PO Box 76, Epping, NSW 2121, Australia}
\altaffiltext{4}{Jet Propulsion Laboratory, 4800 Oak Grove Dr.,
	Pasadena, CA 91109, USA}
\altaffiltext{5}{Dept. of Physics, University of Bristol, Tyndall Ave.,
	Bristol BS8 1TL, UK}
\altaffiltext{6}{Joint Center for Astrophysics and
	Physics Department, Univ. of Maryland,
	Baltimore County, 1000 Hilltop Circle, Baltimore, MD, 21250, USA}
\email{hermanm@space.mit.edu}

\slugcomment{Submitted to Ap J Supplements, \today, hlm}

\begin{abstract}

We present results from {\em Chandra} X-ray imaging
and spectroscopy of a flux-limited sample
of flat spectrum radio-emitting quasars with jet-like
extended structure.  Twelve of twenty quasar jets are
detected in 5 ks ACIS-S exposures.  The quasars without X-ray jets are
not significantly different from those in the
sample with detected jets except
that the extended radio emission is generally fainter.
New radio maps are combined with the X-ray images
in order to elucidate the relation between radio and
X-ray emission in spatially resolved structures.
We find a variety of morphologies, including long
straight jets and bends up to 90\arcdeg.  All X-ray jets
are one-sided although the radio images used for source
selection often show lobes opposite the X-ray jets.
The FR II X-ray jets can all be interpreted as
inverse Compton scattering of cosmic microwave background
photons by electrons in large-scale relativistic jets
although deeper observations are required to test
this interpretation in detail.
Applying this interpretation to the jets as a
population, we find that the jets would be aligned to within
30\arcdeg\ of the line of sight generally, assuming that the
bulk Lorentz factor of the jets is 10.

\end{abstract}


\section{Introduction}

Relativistic jets appear to be a common result of accretion in
both stellar-mass and supermassive black hole systems.  In an active
galactic nucleus (AGN), a jet can transport energy from the sub-parsec scale
central region to a distant hot spot and then
to a radio lobe many times the size of the host
galaxy.  In many active galaxies, the radiative power of the
jet is comparable to that of the nucleus 
and the kinetic energy flux
can be comparable to the bolometric radiative luminosity of the core
\citep{baum,rawlings}.

While jets are commonly detected in radio images of quasars
and radio galaxies, jet detections in the X-ray band were rare
before {\em Chandra}, being limited to the nearest and brightest 
active galaxies, principally Cen~A, M~87, 3C~120, and 3C~273 and hotspots
and jets in a few other objects.
The sensitivity and high-quality imaging possible with
{\em Chandra} now provides more detections
\citep[e.g.][]{schwartz,worrall01,sambruna02,sambruna04}.
Many of the new discoveries are low power (FRI) radio sources,
and BL Lac objects \citep{worrall01,hardcastle01b,pesce01,birkinshaw02}.
In luminous quasars such as 3C 273 and PKS~0637-752, the
jet physics is known to be different from that in low power
sources, so it is important to increase
the number of known X-ray jets in high power sources
and, consequently, our knowledge of their
range of properties.  Two groups are addressing this need
through systematic surveys using {\em Chandra}
observations of radio-selected quasars
(Sambruna, et al.; this work).

Our survey was intended to discover whether radio jets commonly
have X-ray emission that is as bright relative to their radio
fluxes as in the brightest knot in \pks\ \citep{schwartz}.  For
the sources with relatively bright jets, deep exposures
would then provide more detailed measurements of individual
jet components, as has been done, for example, for \pks,
3C 273 \citep{marshall},
M 87 \citep{marshall02,wy02}, and 
Cen A \citep{kraft02,hardcastle03}.
In order to study particle acceleration and energy loss mechanisms
in quasar jets, we need to assess key
issues, such as
whether the particle energy densities are in equipartition
with the local magnetic field energy densities and
whether the jets have high Lorentz factors in bulk motion
oriented close to our line of sight \cite{celotti,tavecchio}.
In this paper, we describe the sample
selection in section 2.  In section 3, we
describe the {\em Chandra} observations and compare
the X-ray maps to newly obtained radio images.  In section
4, we examine the sample properties in the context of
beaming emission models.
We use a cosmology in which $H_o = 65$ km s$^{-1}$ kpc$^{-1}$,
$\Omega_{\rm m} = 0.3$, and $\Omega_\Lambda = 0.7$.

\section{Sample Selection}

We tailored our source selection criteria to generate
a list of objects that have radio jets or nearby bright knots
of suitable angular scale to resolve with {\em Chandra}.
We selected targets from two complete samples for which
radio maps were obtained with $\sim 1$\arcsec\ resolution
at 1-10 GHz: a VLA sample of flat spectrum quasars
($\alpha_{\rm r} < 0.5$,
where $S_{\nu} \propto \nu^{-\alpha}$) with
core flux densities at 5 GHz ($S_5$) $> 1$ Jy and $\delta
> 0$\arcdeg\ \citep{murphy} and an Australia Telescope
Compact Array (ATCA) survey of flat spectrum sources \citep{lovell} from
the Parkes catalog with core flux densities at 2.7 GHz
greater than 0.34 Jy and $\delta < -20$\arcdeg.  The portion of the
sample derived from the ATCA survey is somewhat heterogeneous as
not all sources are optically identified and two have redshifts less than
0.1; these low redshift
sources are likely to be low luminosity FR I galaxies.
Only sources with emission
that was found to be extended over scales $>2$\arcsec\ were
included as candidates.  This selection takes advantage of
the matched radio and X-ray telescope resolutions and of
the excellent point spread function of Chandra: $\lae 5$\%
of the core's power would be outside 2\arcsec, so the
X-ray intensity from the core at the location of 
the jet is expected to be negligible.

The extended radio flux densities
were estimated for each object from previously existing
radio maps of both samples and extrapolated
to 5 GHz, if necessary.  We then estimated the X-ray fluxes
by assuming that the ratio of the 1 keV X-ray flux to the
4.8 GHz flux
is the same as in the jet of \pks, for which $S_{\rm x} = 25$ nJy
and $S_{\rm r} = 0.7$ Jy.  The predicted X-ray count rate is then about
$0.0010 (S_{\rm x}/$nJy$) \approx 0.0357 (S_{\rm r}/$Jy$)$ count/s in the 0.5-7.0 keV band.
We observed sources from the northern list where the predicted counts
would be greater than 30 counts in 5 ks.
For a region of interest that
is about 8\arcsec $\times$ 2.5\arcsec, the limiting count rates
are much larger than the expected background of about 0.0001 count/s.
The southern sample had more sources
with bright extended flux, so the limit was set at 50 counts in 5 ks.
These sources comprise our ``A'' list.

A ``B'' list was also generated from these two
samples.  The selection criterion was based entirely on the
morphology of the extended emission from maps available before
starting the survey: the radio emission was
predominantly one-sided and linear.
This criterion extends our sample to predicted count rate that are
lower than expected for the targets in the A~list.
We include these sources in
our list because we already know that radio surface brightness
is not the only characteristic that sets the X-ray flux.
Knot A1 of 3C 273, for example, has an $S_{\rm x}/S_{\rm r}$
ratio that is 10$\times$ higher than for \pks\ \citep{marshall}.
The morphological selection criterion was
efficient at picking out sources in the A~list, so that many
A-list targets would also qualify in the B~list.

We note that both target lists are drawn from
flux-limited samples based on radio flux without regard
to redshift or optical properties.
Our sample selection is somewhat different from that used by
\cite{sambruna02,sambruna04} who selected sources from a list of known jets
compiled from surveys of the literature.
The completeness of the lists they used is not known.
Six of their targets are also in our sample but of the remaining
eleven, seven do not satisfy at least one of our sample requirements
and four are in the declination gap
(-20\arcdeg\ to 0\arcdeg) between the northern and southern surveys
on which we have based our sample selection.  By selecting our A targets
on the basis of extended flux without regard to morphology, we pick up
more highly bent and amorphous structures than are found in the
lists used by Sambruna et al.  Since
any literature-based sample is biased to the northern hemisphere,
it is not surprising that
many of our new targets are in the southern hemisphere.

The full sample (Table~\ref{tab:sample}) consists of 56 objects, comprised of
9 sources in the A~list that are not in the B~list, 21 sources
that are in both lists, and 26 sources that are only in the
B~list.
Of the 56 sources, 20 were observed as part
of a {\em Chandra} observing program in cycle 3, and are
the subject of this paper.
A few sources in the sample have been observed in
other observing programs and results will be included
in a future paper where we will also include results from
the continuing {\em Chandra} observing program.

\section{Observations and Data Reduction}

\subsection{Imaging results}

The observation dates and
exposure times for the 20 targets observed in {\em Chandra} observing
cycle 3 are given in Table~\ref{tab:observations}.
Events in the 0.5-7.0 keV range were selected for all analysis and
to form the X-ray images, shown in Fig.~\ref{fig:images}.  The location
of the X-ray core in each image
was determined by computing a preliminary centroid by fitting
Gaussians to the 1D histograms obtained from events within 30\arcsec\
of the core.
The final centroid was computed similarly using only events
within 3\arcsec\ of the preliminary centroid.  The observed core count
rate was estimated within a 2.5 pixel (1.23\arcsec) radius aperture and the
(negligible) background was
taken from an annulus from 15 to 20\arcsec\ from the core.
For this size aperture and the typical quasar core spectra,
the enclosed power is about 95\% of the
total power, so we did not apply an aperture correction to the observed
count rates.
Results are shown in table~\ref{tab:coreresults}.

In all but one observation, ACIS was restricted to a 1/4 subarray
with a field of 2\arcmin\ $\times$ 8\arcmin\ and a frametime, $t_{\rm f}$, of 0.8 s.
In one case, 0208-512, a 1/8 subarray (a 2\arcmin\ $\times$ 8\arcmin\ field
of view) was used, for which $t_{\rm f}$ = 0.4 s.
For the brightest sources, pileup, which increases with frametime,
may be a concern.
Pileup diverts the grades of
good events into bad grades, reducing the apparent count rate
and shifting events to higher
energies.
The count rate reduction is greater than 10\% when there are more than about
0.25 counts per frame, or at
about 0.3 cps (0.6 cps) when $t_{\rm f}$ = 0.8 s (0.4 s).
There are three sources for which the {\em observed} core
count rate exceeds the 10\% loss count rate so that the reported count
rate is systematically low by 10\% or more:
1145$-$676, 1828$+$487, 2251$+$158.
We computed independent estimates of the core count
rates using the ACIS readout streak. 
The streak occurs as X-rays are received during the ACIS
parallel frame transfer, which provides 40 $\mu$s exposure in each
pixel along the streak per frame.  For an observation
with a live time of $T$ (yielding $T/t_{\rm f}$ frames), 
a section of the readout streak that is $\theta_{\rm s}$ arcsec long
accumulates an exposure of $t_{\rm s} =
4 \times 10^{-5} T \theta_{\rm s} /(t_{\rm f}  \theta_{\rm x})$ s, where $\theta_{\rm x} =
0.492$\arcsec\ is the angular size of an ACIS pixel.
For these observations, $T \simeq 5$ ks and
$t_{\rm f}$ = 0.8 s (0.4 s), giving
$t_{\rm s} =$ 7.6 s (15.2 s) in a streak segement that
is $\theta_{\rm s}$ = 15\arcsec\ long.
For the three sources in question, the count rates
estimated from the streak are systematically higher, as expected, although
this is not significant for 1145-676 which appears to have a soft
spectral excess (Table \ref{tab:coreresults}).
While the count rate determined from the 2D image is more
precise, it is subject to larger bias due to pileup.
By contrast, the
count rate determined from the streak is more
uncertain but less biased; the streak-based count rates should be
preferred for PKS 1145-676, 1828$+$487, and 2251$+$158.
For 2251$+$158, pileup reduces the observed count rate
by more than a factor of three.

New radio maps were obtained for many of the sources at the ATCA
or the VLA.  These will be reported in detail in a separate paper
(Lovel, et al., in preparation).
These new maps were used for the image overlays in Figure~\ref{fig:images}
(details of the radio contours are provided in Table~\ref{tab:radioCont}),
and were used to determine radio fluxes for the cores and jets.
For the overlays, we assumed that the radio and X-ray cores coincided
in order to register the X-ray and radio maps to better than 0.1\arcsec.
The applied shifts were $\sim 1$\arcsec.

We tested for the existence of an X-ray jet using simple test based
on Poisson statistics.
First, the radio images were used to define the position angles and lengths
of possible X-ray jets.  Most jets are clearly defined as one-sided structures
but in a few ambiguous cases, the pc-scale jets were used to define
the jet direction.
The inner and outer radii, $r_{i}$ and $r_{o}$, are
given in Table~\ref{tab:jetresults} along with the position angle
at which the region was centered.  The width of the rectangle was
3\arcsec\ except for 1030-357, 1655-776 and 2101-490, where the
jets bend substantially, so the
rectangles were widened to 10\arcsec, 10\arcsec, and 5\arcsec, respectively.
Profiles of the radio emission along the jets were extracted and are shown in
fig.~\ref{fig:radioprofiles}.  In order to eliminate the wings of the
core but avoid counterjets, a radio profile was determined
at 90\arcdeg\ to the jet and subtracted.
Next, the X-ray counts in the same rectangular region defined by the radio data
were compared to a similar sized region on the opposite side of
the core for the test, assuming a Poisson probability distribution.
We set the critical probability for detection of an X-ray jet
to 0.0025, which yields a 5\% chance that there might be one
false detection in this set of 20 sources.
Histograms of the X-ray emission along the jets are shown in
Fig.~\ref{fig:xrayprofiles}.  The jet and counter-jet position angles
are compared, providing a qualitative view of the X-ray emission along
the jets.  In no case is a counter-jet apparent in the X-ray images.
X-ray emission that is symmetrically distributed about a quasar core
is ignored because it is either associated with the {\em Chandra}
point response function (so it is more clear in bright sources such
as 2251$+$158) or it is could be an indication of hot gas in the
potential well of a galaxy or small cluster of galaxies.

Jet X-ray fluxes (table~\ref{tab:jetresults})
were computed from count rates using 1000 nJy per count/s.  This
conversion is accurate to about 10\% for typical power law spectra.
The spectral index from radio to X-ray is computed using $\alpha_{rx} = 
\log(S_{\rm x}/S_{\rm r}) / \log(\nu_{\rm x}/\nu_{\rm r})$, where $\nu_{\rm x} = 2.42 \times 10^{17}$ Hz
and $\nu_{\rm r}$ depends on the map used.

\subsection{Core Spectral Fits}

The X-ray spectrum of the nucleus of each source was measured using an
extraction circle of radius 2.5 pixels (1.23 arcsec), with background
measured from source-free regions from the same observation and the
same CCD.  Spectral fitting was performed in the energy band 0.5--7 keV;
the low energy limit was set high enough
to reduce any uncertainties in the ACIS contamination correction
to an unimportant level.
In each case the data were fitted with a power law with excess intrinsic
absorption.
The power-law slope, $\Gamma_{\rm x}$, is the photon spectral index, and thus
is $\alpha_{\rm x} + 1$, where $\alpha_{\rm x}$ is the energy spectral index more
commonly used in radio astronomy and in the inter-comparison of
multiwavelength data.
The X-ray results are given in Table~\ref{tab:coreresults}.

The addition of a narrow K$\alpha$
fluorescence line from neutral Fe at a (fixed) rest energy of 6.4 keV and
(fitted) equivalent width 220 eV improved
the fit significantly in only one quasar: 1202-262.  The detection rate
for this line is consistent with the findings by
\cite{gambilletal03} using the Sambruna et al.\ sample, who find it in three
of 16 quasars observed with {\it Chandra}, particularly given
that their observations were of twice the
duration of ours and their quasar X-ray
cores are typically somewhat brighter than those
we observed.  Only two of our objects, 1828$+$487 and (particularly)
2251$+$158 showed the effects of pileup in the spectra, so for these sources we used the
pile-up model in XSPEC \citep{davis01}.  Spectra of both these objects (the brightest
of our sample) have been measured by earlier X-ray missions.
Our result for the photon index
of 1828$+$487 (3C 380) is in good agreement with the value
of $1.58 \pm 0.13$
($1\sigma$ uncertainty) found over 0.2--2 keV by \cite{prieto96} with
ROSAT.  For 2251+158 (3C 454.3), our measurement is somewhat
lower than the ROSAT results of $1.62 \pm 0.04$
\citep{fossatietal98}
or $1.73 \pm 0.1$ \citep{prieto96}, but in good
agreement with BeppoSax results  of $1.38^{+1.26}_{-1.44}$ (90\%
uncertainty) over a more closely
corresponding energy band \citep{tavecchio02}.
A third object also observed
previously, 0208-512, is reported from
ASCA data to have a spectral index of $1.66^{+0.1}_{-0.09}$
\citep{kuboetal98} or $1.69 \pm 0.04$
\citep{reevesturner00}, and from BeppoSAX to have
$1.7^{+1.96}_{-1.62}$ \citep[90\% uncertainty, with an absorption
feature at 0.6 keV of which there is no evidence in the Chandra
data][]{tavecchio02} in agreement with our value.
Also observed with ROSAT in a pointed observation, it has a reported
spectral index in
the soft X-ray band of $2.04 \pm 0.04$ \citep{fossatietal98}, suggesting
either variability or a soft
excess (see also Sambruna 1997).  Soft excesses are not uncommon in quasar cores, and
indeed the {\it Chandra\/} data suggest one is present in 1145-676
(Table~\ref{tab:coreresults}).

Core radio and X-ray flux densities are plotted against each other
in Fig.~\ref{fig:corefluxes}.
The radio flux densities were measured from
the maps shown in Fig.~\ref{fig:images} and then adjusted to 5 GHz
assuming $\alpha_{\rm r} = 0.25$.
The X-ray flux densities are estimated at 1 keV ($2.42 \times 10^{17}$ Hz).
The radio and X-ray maps were not obtained at the same time, so there will
be additional scatter in Fig.~\ref{fig:corefluxes} due to variability.
The spectral indices between the radio and X-ray bands ($\alpha_{rx}$) are
centered on a value of 0.5 and the targets in subsample A are not
distinguishable from those in subsample B, indicating that the
morphology selection does not
pick out targets with preferentially dim X-ray cores even though the extended
radio fluxes are somewhat smaller.  Similarly, jet detections are found without
bias regarding either this spectral index
or the radio or X-ray core flux densities.

The X-ray photon indices, $\Gamma_{\rm x}$, are
collected in Figure~\ref{fig:xrayspectra}.
We follow practice dating from the {\it Einstein\/}
Observatory period of assuming
that the underlying spectral-index distribution is Gaussian,
and maximize the likelihood to find the best-fit
underlying mean and dispersion, with 90\% joint-confidence 
uncertainties, are $\bar\Gamma =
1.59^{+0.07}_{-0.09}$ and $\sigma =
0.13^{+0.11}_{-0.05}$.  The results are consistent with
early X-ray measurements of the cores, and support the conclusion that
flat-radio-spectrum quasars have flatter average X-ray spectra than
radio-quiet quasars, consistent with a beamed radio-related
component contributing to the X-ray emission from radio-loud quasars
(see Worrall [1990] for a review). 
\citet{gambilletal03} found similar values of
$\bar\Gamma = 1.66$ and $\sigma = 0.23$, with our somewhat smaller value of
$\sigma$ possibly implying that our sample is more homogeneous in its
selection.  We find no obvious difference between our subsamples with
and without jets.

\subsection{Notes on Individual Sources}

In this section, we present qualitative descriptions of the X-ray and
radio morphologies shown in Figure~\ref{fig:images} and describe the
orientations of the polarizations or any pc-scale jets.  Profiles of the 
radio and X-ray emission along the jets are given in figures~\ref{fig:radioprofiles}
and \ref{fig:xrayprofiles}, respectively. 

\subsubsection*{0208$-$512}

Our ATCA image of 0208$-$512 shows a $\sim$4\arcsec\ long jet at a
position angle of $-129$\arcdeg, which is similar to the position
angle of the milliarcsecond-scale radio jet \citep{2002ApJS..141..311T},
indicating a straight jet from pc to kpc scales. X-ray emission is
also seen in the arcsec-scale jet but fades drastically after a
90\arcdeg\ bend in the radio jet (Fig.~\ref{fig:images}a). The radio
polarization electric vectors in the 4\arcsec\ jet are orthogonal to
the jet direction but no polarized emission is detected in the jet
after the bend.  A more detailed analysis is presented by
\cite{schwartz04}.

\subsubsection*{0229$+$131}

The radio emission from 0229$+$131 shows a bright inner jet to the
NE which bends to the east and weakens (Fig.~\ref{fig:images}b). There is also a weak
westerly jet.
VLBA images (e.g \cite{2000ApJS..131...95F}) show a milliarcsec jet which
initially points to the NE but then bends to the east $\sim$4~mas from
the core.
We detect no associated extended X-ray emission.

\subsubsection*{0413$-$210}

The VLA image of 0413$-$210 reveals a two-sided radio
jet. The brighter jet is to the SE for which there is a corresponding
X-ray jet (Fig.~\ref{fig:images}c) and the radio polarization data show
electric vectors orthogonal to the jet direction. This source has also
been observed at milliarcsec resolution as part of the VLBA calibrator
survey \citep{2002ApJS..141...13B} and shows a single-sided jet at a
similar position angle to its arcsec-scale counterpart to the SE.

\subsubsection*{0745$+$241}

0745$+$241 shows a bright radio core with weak extended lobes.  VLBA
observations by \cite{2000ApJS..131...95F} show a one-sided jet to the
NW indicating that the jet bends by $\sim$40\arcdeg\ between
the milliarcsec- and arcsec-scale structure.
There are four photons (in
a 0.7\arcsec\ radius circle) associated with the western lobe at
about 7\arcsec\ from the core, which constitutes
a detection at about the 3.6$\sigma$ level, where the background was
determined in an annulus from 6.0 to 8.8\arcsec\ (Fig.~\ref{fig:images}d).
Otherwise, we detect no significant extended X-ray emission related
to the radio emission.

\subsubsection*{0858$-$771}

Our ATCA image of 0858$-$771 shows a two-sided jet (Fig.~\ref{fig:images}e). No
linearly polarized flux was detected in the core with a 3-sigma detection
limit of 2~mJy. In the southern jet the electric vectors are
orthogonal to the jet direction while in the northern jet the
polarization structure is more complex. No polarized flux is detected
along the weak inner part of the northern jet but the lobe is
polarized by up to 12\%. On the western side of the lobe, where the
inner jet terminates, the electric vectors are parallel to the jet
direction. However on the eastern side of the lobe the E-vectors have
rotated by 90\arcdeg. It is at the termination of the inner north jet
where we find a few X-ray photons that do not comprise a
significant detection of the jet or its termination.
No VLBI data are available for this source to
give an idea of the initial directions of the jets.
We detect no associated extended X-ray emission.

\subsubsection*{0903$-$573}

The radio emission from 0903$-$573 reveals a one-sided
arcsec-scale jet to the NE of the core and low-level extended
emission (Fig.~\ref{fig:images}f). There is X-ray emission
associated with the radio jet. The
radio polarization E-vectors along the jet are at position angles
between 70\arcdeg\ and 80\arcdeg\ compared to a jet
direction of $\sim$33\arcdeg. The milliarcsec-scale structure of
this source is presently unknown.

\subsubsection*{0920$-$397}

There is a strong X-ray detection in the southern jet of
0920$-$397 (Fig.~\ref{fig:images}g).  The X-ray brightness drops monotonically along the
5\arcsec\ long inner jet and there is perhaps an X-ray detection
coincident with the southern radio lobe
(Fig.~\ref{fig:images}g).  Data in the United States Naval Observatory
Radio Reference Frame Image Database
reveal a milliarcsec-scale jet to the south indicating that the
southern jet remains straight from pc to kpc scales. No radio
polarization was detected in the jets of this object.
A more detailed analysis is presented by \cite{schwartz04}.

\subsubsection*{1030$-$357}

X-ray emission is marginally detected from the inner 8\arcsec\ of the
jet that has a slight bend towards the southeasterly of
two bright radio features about 13\arcsec\ from the core
(Fig.~\ref{fig:images}h).
Two X-ray bright, resolved knots are associated with the bright
features, which may be interpreted as
two 90\arcdeg\ bends in the jet. The radio
polarization electric field vectors are orthogonal to the jet
direction at least as far as the first knot.  New VLBI images show
a jet about 0.002\arcsec\ long at nearly the same position angle
as the 10\arcsec\ scale jet \citep{ojha}.  A more detailed analysis
of the X-ray and radio data is presented by \cite{schwartz04}.

\subsubsection*{1046$-$409}

The radio structure of this
source is somewhat confusing, with two knots to the
northwest, a knot 4\arcsec\ to the southeast at the beginning of an
arc curving 90\arcdeg, and amorphous emission 20\arcsec\ in diameter
centered on the core (Fig.~\ref{fig:images}i). X-ray emission is
detected at the beginning of the arc to the southeast, which we interpret
as the brightest part of a curved jet. Radio polarization
E-vectors are orthogonal to the jet direction out to the first knot to
the SE but are parallel to the jet direction at the end of the
arc. VLBA observations by \cite{2000ApJS..131...95F} show a one-sided
milliarcsec jet at a position angle similar to that of the initial
direction of the southeast arcsec-scale jet. Once again the change in
radio jet direction and polarization appears linked to the termination
of the X-ray jet.

\subsubsection*{1145$-$676}

The one-sided jet curves slightly, ending in a resolved structure perpendicular
to the jet (Fig.~\ref{fig:images}j).  The western radio lobe is not
symmetrically placed opposite the jet.
X-ray emission is not formally detected from the jet when the entire jet
length, 40\arcsec, is used.
Defining the selection region to be only 5\arcsec\ long, however, 
does lead to the X-ray emission becoming significant.
The radio polarization E-vectors are orthogonal
to the direction of the inner 5\arcsec\ jet.
 
\subsubsection*{1202$-$262}

This target has the brightest X-ray jet in our sample,
with a $\sim$30\arcdeg\ bend that follows the
radio image until the radio jet takes a $\sim$90\arcdeg\ bend
(Fig.~\ref{fig:images}k).  This jet is remarkable in that the flux
in the jet is 10\% of the power in the core (after accounting for
pileup) while the highest jet/core ratio previously obtained was 7\%
for \pks\ \citep{schwartz}.  The radio data reveal polarization
E-vectors perpendicular to the jet direction along the northern jet as
far as the lobe 6\arcsec\ from the core where the position angle has
rotated by 90 degrees. On the milli-arcsecond scale there is also a jet
aligned with the initial direction of the northern arcsec-scale jet
\citep{2000ApJS..131...95F}.   The southern radio lobe is well resolved and
symmetrically positioned opposite the extended emission to the
northeast.  A more detailed analysis is presented by
\cite{schwartz04}.

\subsubsection*{1258$-$321 = ESO 443$-$G024}

The source has a low redshift ($z = 0.01704$) and extended radio morphology
that is concentrated close to the core, indicating that this source
is a low-power FR I radio source (Fig.~\ref{fig:images}l).
We do not detect X-ray emission associated with the radio jets.
There may be X-ray emission from an extended halo of hot gas
about 10\arcsec\ across but we have not specifically tested this
hypothesis.

\subsubsection*{1343$-$601 = Cen B}

This galaxy also has a low redshift ($z = 0.01292$) but the radio
morphology is clearly one-sided and shows a narrow jet with a few
knots (Fig.~\ref{fig:images}m).  A knot 5\arcsec\ from the core is
easily detected in the X-ray image as well as another region 
8\arcsec\ from the core. The polarized component of the radio emission
shows 
E-vectors approximately perpendicular to the jet direction but shows
some deviations, particularly at the brightest part of the X-ray jet.
There is well-resolved X-ray emission of 0.017 $\pm$ 0.002
count/s in a region spanning 10-70\arcsec\ from the core, upstream
of and coincident with the diffuse lobe.  New VLBI images show
a jet about 0.020\arcsec\ long at nearly the same position angle
as the 10\arcsec\ scale jet \citep{ojha}.

\subsubsection*{1424$-$418}

1424$-$418 is one of the most compact radio sources in our sample.
The radio data show a jet extending 1\arcsec\ to the west of a
compact core before becoming diffuse (Fig.~\ref{fig:images}n).
We detect no associated extended X-ray emission.

\subsubsection*{1655$+$077}

This quasar ($z=0.621$) shows a compact radio core, a short jet
extending
to the SE and a jet extending 2.5-3\arcsec\ to the NW before bending
90\arcdeg\ to the SW (Fig.~\ref{fig:images}o).  VLBA data show knots along
a 0.007\arcsec\ long jet to the NW \citep{2000ApJS..131...95F}.
No X-ray emission was detected other than in the core.

\subsubsection*{1655$-$776}

This galaxy ($z=0.0944$) shows a compact radio core, low surface
brightness lobes and a hot-spot in the eastern lobe (Fig.~\ref{fig:images}p).
With the jet length and width parameters from Table~\ref{tab:jetresults},
we detect no X-ray emission other
than in the core.  The image seems to show a weak X-ray jet to the ENE
that is only $\sim$ 5\arcsec\ long but this emission is only marginally
significant with $P_{jet} = 0.080$.

\subsubsection*{1828$+$487 $=$ 3C 380}

The radio
polarization electric vectors are parallel to the direction of the
2\arcsec-long inner jet, oriented at a position angle of
$-40$\arcdeg.  A pc-scale superluminal
radio jet \citep{1998MNRAS.294..327P} is aligned with this arcsec-scale
jet.
The {\em Chandra} point spread function is just small enough to
provide a detection of a knot in the inner jet about 1.8\arcsec\ from
the core coincident
with the inner radio jet (Fig.~\ref{fig:images}q).  See also the
X-ray and radio profiles
(figs.~\ref{fig:xrayprofiles} and \ref{fig:radioprofiles}).

\subsubsection*{2052$-$474}

This redshift 1.489 quasar has a bright radio core and a two-sided
arcsec-scale radio jet (Fig.~\ref{fig:images}r).
VLBA data on this object have not revealed the
presence of a milliarcsec-scale jet \citep{1998ApJ...500..673T}.
We detect no extended X-ray emission.

\subsubsection*{2101$-$490}

2101$-$490 shows remarkably similar radio morphology to
\pks (Fig.~\ref{fig:images}s).
There is a long straight jet with three hotspots, after which the
jet bends and terminates in a bright lobe. The polarisation E-vectors remain
orthogonal to the jet direction as far as the lobe where they are
parallel. Several X-ray knots are also seen along the jet but unlike
\pks\ they appear to be associated with areas of weak radio emission,
and a knot exists past the bend.
We have assigned a tentative redshift of 1.04 to
2101$-$490 based on a spectrum obtained at the Magellan telescope
\citep{gm04}.
New VLBI images show
a jet about 0.004\arcsec\ long at nearly the same position angle
as inner part of the 12\arcsec\ long jet \citep{ojha}.

\subsubsection*{2251$+$158 $=$ 3C 454.3}

This redshift 0.859 quasar
has a bright radio core and a one-sided
arcsec-scale radio jet that was not intrinsically bright enough
to be in the A subsample.
A knot at the end of the jet is easily
detected in the X-ray image (Fig.~\ref{fig:images}t).
A pc-scale jet with components starting at a position angle -29\arcdeg\ and
ending at a position angle of -80\arcdeg\ was detected in VLBI
observations \citep{lobanov00}.
VLBA data on this object shows a jet
that curves from a position angle of -115\arcdeg\ to -60\arcdeg\ where
it  begins to align with the larger scale jet that is detected in
X-rays.

\section{Discussion}

The null hypothesis that bears testing with these data is that
the jet is highly relativistic and aligned close to the line of sight
so that the
X-ray emission results from inverse Compton scattering of cosmic microwave
background (IC-CMB) photons off jet electrons \cite{celotti,tavecchio}.
We have several lines of evidence that suggest that the jets in our sample
are consistent with this interpretation.

\subsection{Detection Statistics}

We detect 12 new X-ray emitting jets among the 20 targets observed.
Of these detections, 9 were in the A subsample of 10 sources, while
only 3 were in just the B subsample.  If detections were equally likely
in both B and A samples, then the {\it a priori} probability that there
would be $<4$ B detections would be 7.3\%, so the hypothesis that the
morphology selection is just as good as a flux selection remains
tenable.
Completing the {\em Chandra} imaging of these quasars may
provide enough data to reject this hypothesis.
Of the full sample of 56 sources, 19 of the A subsample have been observed using
{\em Chandra} and 16 of these have jets that are detected in short
exposures, for an 84\% detection rate; the typical flux densities
of detected jets are greater than 2 nJy.  This detection rate is slightly higher
than that obtained by \cite{sambruna04}.  The jet detection rate for
the B-only subsample is not as high.

Fig.~\ref{fig:alpharx} shows that the B targets were generally
fainter in the radio band, so that the limits on $\alpha_{rx}$ were higher
(i.e. fainter X-ray fluxes detected relative to the radio fluxes) than for
the A targets.  The B targets are mostly one-sided by selection, so,
if sufficiently more B targets are detected in future observations to
reject the hypothesis of similar A and B detection rates, then we
would conclude that morphological selection yields sources with brighter
X-ray jets, as might be expected if these are more closely aligned to the
line of sight and are Doppler boosted.

\subsection{Modeling the X-ray Emission}

For a one population
synchrotron model to extend from the radio to the X-ray band, the
spectral index $\alpha_{rx}$ would have to be comparable to that in the
radio band.  If the radio spectral index is $\alpha$, then synchrotron
models with a single power law distribution of electrons
can be feasible if $\alpha_{rx} \geqslant \alpha$ but then the optical flux would
also be bright, so that $\alpha_{rx} \geqslant \alpha_{ro} \geqslant \alpha$.  Our
program of imaging using the Magellan telescope \citep{gm04} places
stringent limits on the optical emission from most X-ray emitting jets
in the southern hemisphere,
ruling out the single population synchrotron model.

Using the IC-CMB formulation given by \cite{hk02}, we may obtain a crude
estimate of the angle to the line of sight with some physical assumptions
such as requiring that the magnetic field can be determined from the minimum
energy argument.
\cite{hk02} determined an expression for a combination of the bulk $\Gamma$
and the angle to the line of sight, $\theta$, in terms of a few quantities
derived from observables.  Following their approach, for all jets in our
sample, we first define $B_1$, the
spatially averaged, minimum energy magnetic field in a jet in the case
where there is no Doppler boosting ($\Gamma = 1$)

\begin{equation}
\label{eq:b1}
B_1 = \bigg [ \frac{18.85 C_{12} (1+k) L_{\rm s}}{f V} \bigg ]^{2/7}  {\rm G}
\end{equation}

\noindent
where $\alpha = 0.8$, $C_{12}$ is a weak
function of $\alpha$ and is $5.7\times10^7$
for $\alpha = 0.8$, the filling factor is $f = 1$,
the baryon energy fraction parameter $k = 0$, and the
synchrotron spectrum (with index $\alpha$)
used to determine the synchrotron luminosity, $L_{\rm s}$,
(in erg s$^{-1}$) extends from $\nu_1 = 5 \times 10^6$ Hz
to $\nu_2 = 10^{15}$ Hz.  The emission volume (in cm$^{3}$) is
estimated using the length of
the jet as $\theta_o - \theta_i$ and assuming a cylindrical cross section given
by the diameter associated with the {\em Chandra} FWHM: 0.75\arcsec.
The filling factor is somewhat too large in general as most
jets consist of a series of unresolved knots.  The quantity $B_1$ increases
as $f^{-2/7}$, so $B_1$ would increase by $\times 1.6$ if $f$ is as small as 0.2.
The resultant magnetic fields are given in table~\ref{tab:beaming}.

Next, we compute $R_1$, the ratio of the sychrotron to inverse Compton
luminosities.  As long as both the radio and X-ray frequencies are far from
the endpoints of the synchrotron and inverse Compton spectral breaks, then

\begin{equation}
\label{eq:ratio1}
R_1 \equiv
\frac{S_{\rm x} (\nu/\nu_{\rm x})^{-\alpha}}{S_{\rm r} (\nu/\nu_{\rm r})^{-\alpha}} =
\frac{S_{\rm x} \nu_{\rm x}^\alpha}{S_{\rm r} \nu_{\rm r}^\alpha} =
	(\nu_{\rm x}/\nu_{\rm r})^{\alpha-\alpha_{rx}}  ,
\end{equation}

\noindent
where $\nu_{\rm r}$ and $\nu_{\rm x}$ are the radio and X-ray frequencies at which the
flux densities $S_{\rm x}$ and $S_{\rm r}$ are observed.  We then compute the quantity

\begin{equation}
\label{eq:kdef}
K = B_1 (a R_1)^{1/(\alpha+1)} (1+z)^{-(\alpha+3)/(\alpha+1)}
	b^{(1-\alpha)/(\alpha+1)}  ,
\end{equation}

\noindent
where $a = 9.947 \times 10^{10}$ G$^{-2}$ and
$b = 38080$ G are constants and
$B_1$ is expressed in G, so K is dimensionless.
For $\Gamma \gg 1.5$, then
$\Gamma^{\alpha+1} \gg \frac{\Gamma^{\alpha-1}}{4}$, so we may use 
eq.\ A24 from \cite{hk02}:

\begin{equation}
\label{eq:beaming}
K = \Gamma \delta (1+\mu_j^\prime) = \frac{1 - \beta + \mu - \mu \beta}
	{(1-\mu \beta)^{2}}  ,
\end{equation}

\noindent
where we have substituted the equation for the cosine of the angle to the
line of sight in the jet frame, $\mu_j^\prime$, with the angle transformation
formula $\mu^\prime = \frac{\mu - \beta}{1-\mu \beta}$ and the formula for the
Doppler factor $\delta = [\Gamma (1-\mu \beta)]^{-1}$ and rearranged.  We solve
eq.~\ref{eq:beaming} for $\mu$ to obtain the angle to the line of sight for any
value of $K$ and an allowable value of $\beta$:

\begin{equation}
\label{eq:mu}
\mu = \frac{1-\beta +2 K \beta - (1-2 \beta + 4 K \beta + \beta^2 
	- 4 K \beta^3)^{1/2}}{2 K \beta^2}  ,
\end{equation}

By assuming that all the observed quasar jets have the same $\Gamma =
1/(1-\beta^2)^{1/2} = 10$, then we find that the jets have angles $\theta$
to the line of sight as given in table~\ref{tab:beaming}.  Limits are given for
jets that are not detected in X-rays.  The angles are rather robust to many
of the assumed physical parameter.  For example, $\theta$
decreases 10\% if $B_1$ increases 60\% ($B_1$ includes the estimate of
the volume and filling factor, see eq.~\ref{eq:b1}), increases 15\% if
$\nu_2$ decreases to $10^{11}$ Hz, and drops 15\% if $\alpha$ is decreased
to 0.70.  The angles range from 9\arcdeg\ to over 40\arcdeg.  The values for
the FR I radio galaxy, 1343$-$601, are not realistic because the jet is
not likely to be relativistic, so eq.~\ref{eq:beaming} wouldn't apply.
These angles are meant only to be indicative for the remaining
jets; better models would be obtained
by fitting the spectra of individual knots within the jets.
In this model, where $\gamma = 10$ is assumed, the angles determined for the
jets are generally small, often less than 20\arcdeg.
Small angles to the line of sight are expected due to the sample selection
(e.g., most are flat-spectrum, core-dominated radio sources).
Some of the cores show relativistic jets in VLBI images; these would be
apparently brighter than quasars with jets perpendicular to the line of sight
and more likely to populate the parent flux-limited
sample.  The jet angle distribution
depends on the assumed value of $\Gamma$, which may also be distributed, so
we have chosen a value which is close to the median of values
determined from VLBI observations of quasar cores.

Another potentially important possible source
of variance is in the detailed shape of the electron energy distribution.
Because the radio band results from radiation by a different set of particles
than produces the X-ray emission in this model, the derived beaming parameters
would change if
electron number densities do not follow the power law model.
We can estimate the radio frequency, $\nu_{\rm s}$, where the electrons
dominate the synchrotron radiation spectrum that
upscatter CMB photons into the X-ray band to energies
of $h \nu_{\rm IC} = $1 keV.
Starting from eq.~B2 of \cite{hk02}:

\begin{equation}
\nu_{\rm s} = \frac{\nu_{\rm IC} B^\prime}{b \Gamma (1+z) (1+\mu_j^\prime)}
\end{equation}

\noindent
where $B^\prime = B/\delta$ is the magnetic field in the rest frame.
Then, using eqs.~\ref{eq:kdef} and \ref{eq:beaming}, we find a simple relation
between the ratio of the synchrotron and IC frequencies, depending only on
the luminosity ratio, $R_1$:

\begin{equation}
\frac{\nu_{\rm s}}{\nu_{\rm IC}} = \bigg[ \frac{ (1+z)^2 }{a b^2 R_1}
	\bigg]^{\frac{1}{\alpha+1}}
\end{equation}

\noindent
Values of $\nu_{\rm s}$ were determined for each source and are given in
Table~\ref{tab:beaming}.  All values are between 5 and 35 MHz, which is
well below the frequncies of the radio maps we use to measure the knots
(but above the synchrotron cutoff used to determine the synchrotron luminosity).

In the IC-CMB model, the X-ray flux increases as the local energy density
in the CMB increases with redshift.  From 
eq.~\ref{eq:beaming}, we expect $R_1 \propto (1+z)^{3+\alpha}$.
In Fig.~\ref{fig:alpharx-z}, we
show the measurements and limits on $\alpha_{rx}$ as a function of $(1+z)$,
where we set $z=1$ for the source without a measured redshift.
If the IC-CMB model were the dominant mechanism for determining the X-ray
flux, one might expect to find smaller values of $\alpha_{rx}$ at high $z$.
No such trend is indicated in the data, which might just indicate that other
factors dominate the distribution of $\alpha_{rx}$ such as angles to the line
of sight and bulk $\Gamma$.  The distribution of $\alpha_{rx}$ at any given
redshift can be quite broad, corresponding to a range in $R_1$ of
over a factor of 40 near $z = 0.8$.

There were no X-rays detected from counter-jets.  This result is consistent
with the IC-CMB hypothesis, where X-ray emission would only be observed from the
approaching jet due to Compton scattering into the line of sight.  Consider
the ratio, $\xi$, of $R_1$ for a receding jet to that of an oppositely pointed,
similar velocity approaching jet.  This ratio can be determined from
equations~\ref{eq:kdef} and \ref{eq:beaming} by reversing the sign of $\mu$:

\begin{equation}
\xi \equiv \frac{R_1^+}{R_1^-} = \bigg[ \frac{1 - \mu}{1 + \mu}
	\bigg( \frac{1-\mu \beta}{1+\mu \beta} \bigg)^{2} \bigg]^{1+\alpha}
\end{equation}

\noindent
The quantity $\mu$ is defined to be positive for the approaching jet
and using $\beta < 1$, it is easy to show that that $\xi < 1$ for all
choices of $\mu$ and $\beta$.  Thus, the inverse Compton
X-ray emission from a
relativistic counterjet will always be weaker relative to the
radio emission than for the approaching jet.

Detailed analysis of several of the brighter quasar
jets shows that even the meager data for
some of the knots already require large beaming factors and
small angles to the line of sight.
\cite{schwartz03} and \cite{schwartz04} used the
spatially resolved radio/X-ray SEDs of
four sources from this sample to measure magnetic
fields and relativistic beaming factors as a function of distance
along the jet, and inferred the kinetic powers and radiative
efficiencies of the jets.  The minimum energy magnetic fields are
estimated from the radio fluxes and approximate values for the angular diameters
of the knots.
As expected, \cite{schwartz04} found that the high X-ray fluxes required
can be explained reasonably well in the IC-CMB model.
The rest frame magnetic fields of the
different jets were of order 10 to 20 $\mu$G; our values are generally
smaller, 4-14 $\mu$G,
due to the larger regions used.  The measured beaming
factors of 2.5 to 10.6 indicate that the jet axes must be within 5 to
25\arcdeg\ of our line of sight, while
we find angles of 15 to 28\arcdeg\ for these same
sources using the total jet flux densities.
The kinetic powers are of
the same order as the core luminosities for these jets, and their
radiative efficiencies are $\sim$ 1\% or less.
Deeper observations will improve the analysis of individual knots in
these jets and provide an opportunity to test the IC-CMB model
better by providing a measure of the X-ray spectra and perhaps
additional knot detections.

\acknowledgments

We thank Brendan Miller, Jeff Bridgham, Chris Copperwheat,
and Martin Terry for their contributions
to the analysis at an early stage.
Support for this work was provided by the National Aeronautics and Space
Administration (NASA) through Chandra Award Number NAG 5-10032 issued by the
Chandra X-Ray Center (CXC), which is operated by the Smithsonian
Astrophysical Observatory for and on behalf of NASA under contract NAS
8-39073.  HLM was supported under NASA contract SAO SV1-61010 for the CXC.
DAS was partially supported by 
Chandra grant G02-3151C to SAO from the CXC.
JMG was supported under Chandra grant GO2-3151A to MIT from the CXC.
This research has made
use of the United States Naval Observatory Radio Reference
Frame Image Database.  The Australia Telescope Compact
Array is part of the Australia Telescope which is funded by the
Commonwealth of Australia for operation as a National Facility managed
by CSIRO.  This research has made use of the NASA/IPAC Extragalactic
Database (NED) which is operated by the Jet Propulsion Laboratory,
California Institute of Technology, under contract with the
National Aeronautics and Space Administration.

\clearpage

\begin{figure}[htp]
  \plotone{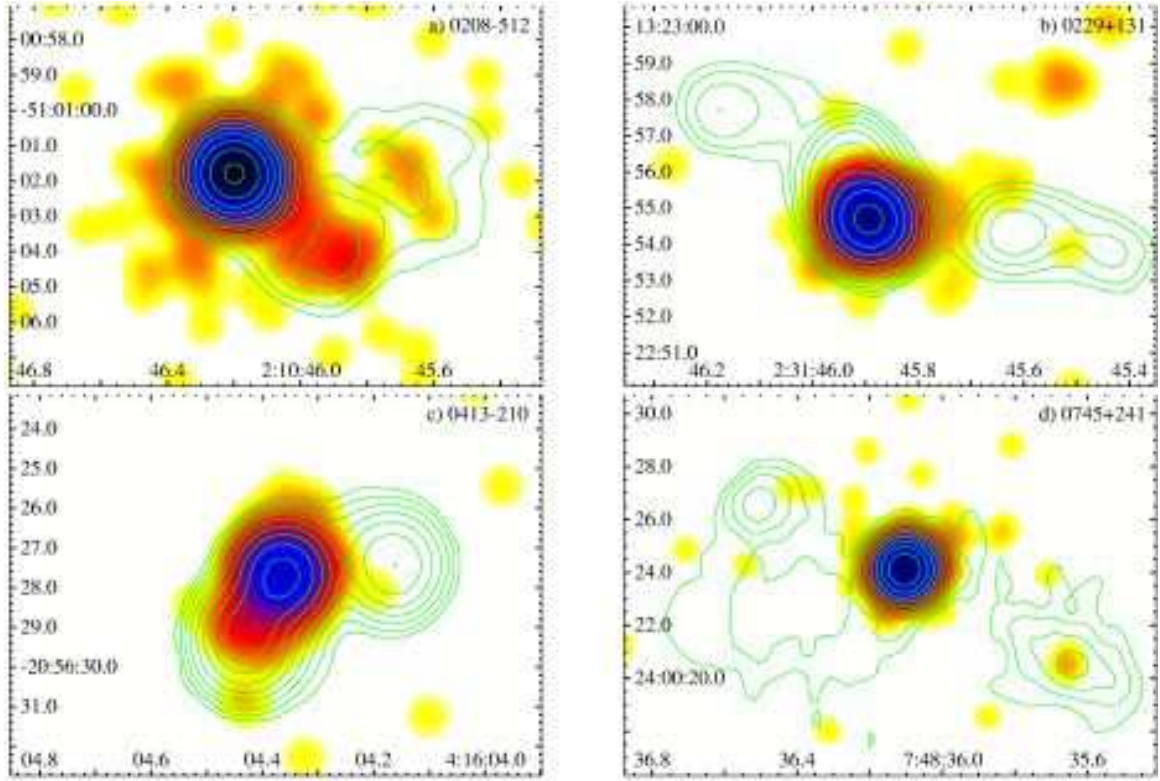}
  \caption{X-ray images obtained with the Chandra X-ray Observatory,
  overlaid by contours of radio emission obtained at the Australia
  Telescope Compact Array or the Very Large Array (VLA).
  The images appear in the following order:
  {\em a)} 0208$-$512,
  {\em b)} 0229$+$131,
  {\em c)} 0413$-$210,
  {\em d)} 0745$+$241,
  {\em e)} 0858$-$771,
  {\em f)} 0903$-$573,
  {\em g)} 0920$-$397,
  {\em h)} 1030$-$357,
  {\em i)} 1046$-$409,
  {\em j)} 1145$-$676,
  {\em k)} 1202$-$262,
  {\em l)} 1258$-$321,
  {\em m)} 1343$-$601,
  {\em n)} 1424$-$418,
  {\em o)} 1655$+$077,
  {\em p)} 1655$-$776,
  {\em q)} 1828$+$487,
  {\em r)} 2052$-$474,
  {\em s)} 2101$-$490, and
  {\em t)} 2251$+$158.
  The flux densities increase by $\times 2$ for
  each radio contour, starting at a value of five times the rms noise.
  Two images are shown with extra contours (in red): 
  PKS 1030$-$357 (Fig.~\ref{fig:images}h) includes two contours from
  a lower-frequency radio map (at $5\times$ and $10\times$ the
  rms level) to illustrate larger-scale structures, and
  PKS 1343$-$601 (Fig.~\ref{fig:images}m) adds an extra contour at
  $3\times$ the rms level to emphasize the diffuse structure.
  The synthesized radio Gaussian beamsizes
  are convolved to produce
  circular 1.2\arcsec\ FWHM beams in all images.
  Radio contour details are provided in Table~\ref{tab:radioCont}.
  The X-ray images are convolved
  with Gaussians to match the radio beamsizes and then
  binned at 0.0492\arcsec.  The color scales are the
  same in all images, ranging logarithmically
  from 0.5 counts/beam (yellow) to 2500 counts/beam
  (black).  Notes on individual objects are given in the text.
  } \label{fig:images}
\end{figure}

\addtocounter{figure}{-1}

\begin{figure}[htp]
  \plotone{f1b-72.epsf}
  \caption{continued.}
\end{figure}

\addtocounter{figure}{-1}

\begin{figure}[htp]
  \plotone{f1c-72.epsf}
  \caption{continued.}
\end{figure}

\addtocounter{figure}{-1}

\begin{figure}[htp]
  \plotone{f1d-72.epsf}
  \caption{continued.}
\end{figure}

\addtocounter{figure}{-1}

\begin{figure}[htp]
  \plotone{f1e-72.epsf}
  \caption{continued.}
\end{figure}

\addtocounter{figure}{-1}

\begin{figure}[htp]
  \plotone{f1f-72.epsf}
  \caption{continued.}
\end{figure}

\begin{figure}[htp]
\epsscale{0.8}
\plotone{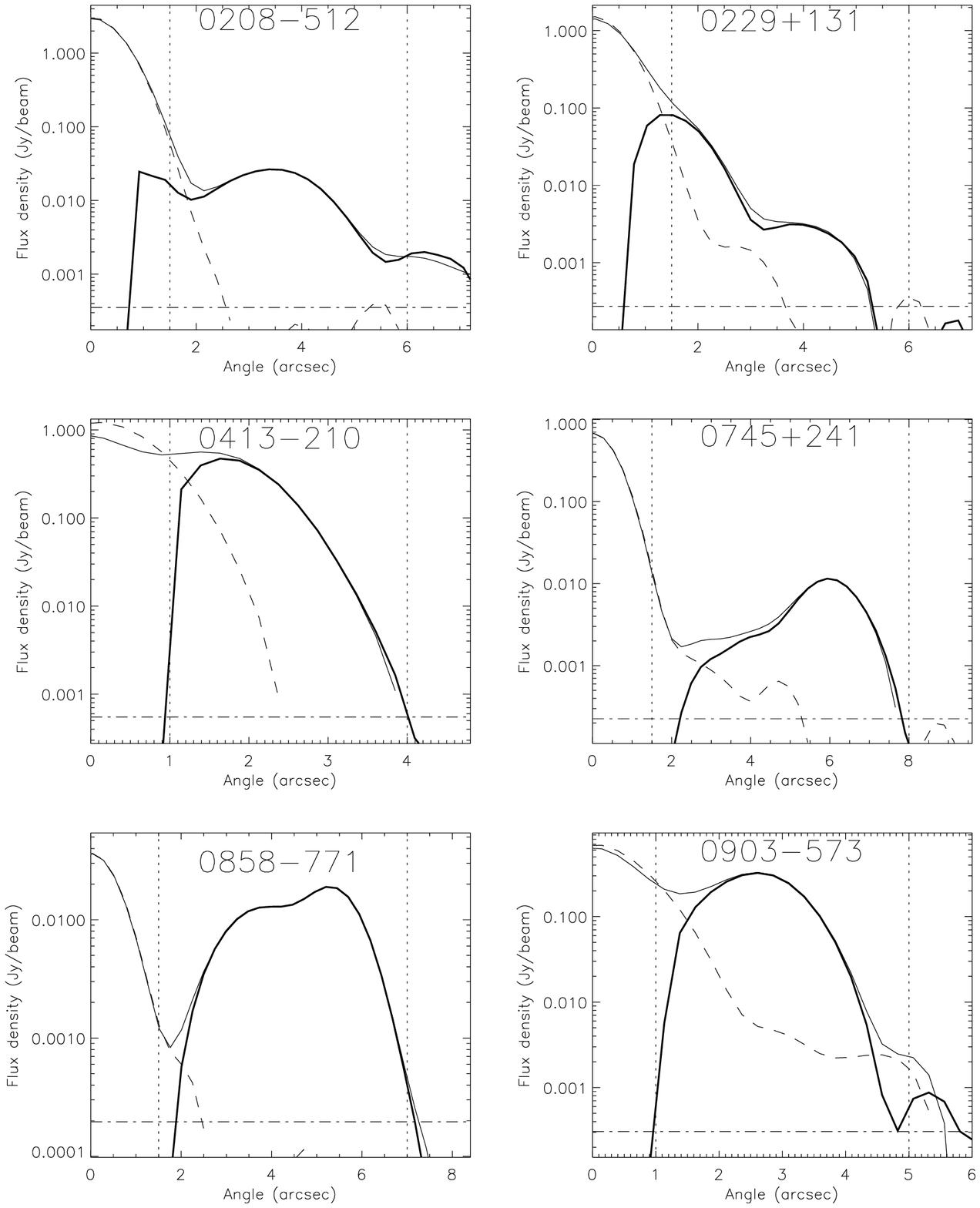}
  \caption{Profiles of the radio images.  The solid, thin
  lines give the profiles along the position angles of the jets, as defined
  in Table~\ref{tab:jetresults} and used for measuring the X-ray profiles.
  The counts are determined in rectangles 3\arcsec\ wide, except for
  except for 1030$-$357 and 2101$-$490, where the jets bend
  substantially, so the rectangles were widened to 10\arcsec\ and
  5\arcsec, respectively.
  The dashed lines give the profiles at
  a position angle 90\arcdeg\ CW from the jet to avoid any counter-jets
  or lobes opposite the jet.
  The solid, bold lines give the difference between the profiles along the
  jet and perpendicular to it, so that the core is effectively nulled and
  the jet flux can be measured as a residual between the vertical dotted lines.
  The horizontal dash-dot lines give the average noise level in the map.
  } \label{fig:radioprofiles}
\end{figure}

\addtocounter{figure}{-1}

\begin{figure}[htp]
\epsscale{1.0}
\plotone{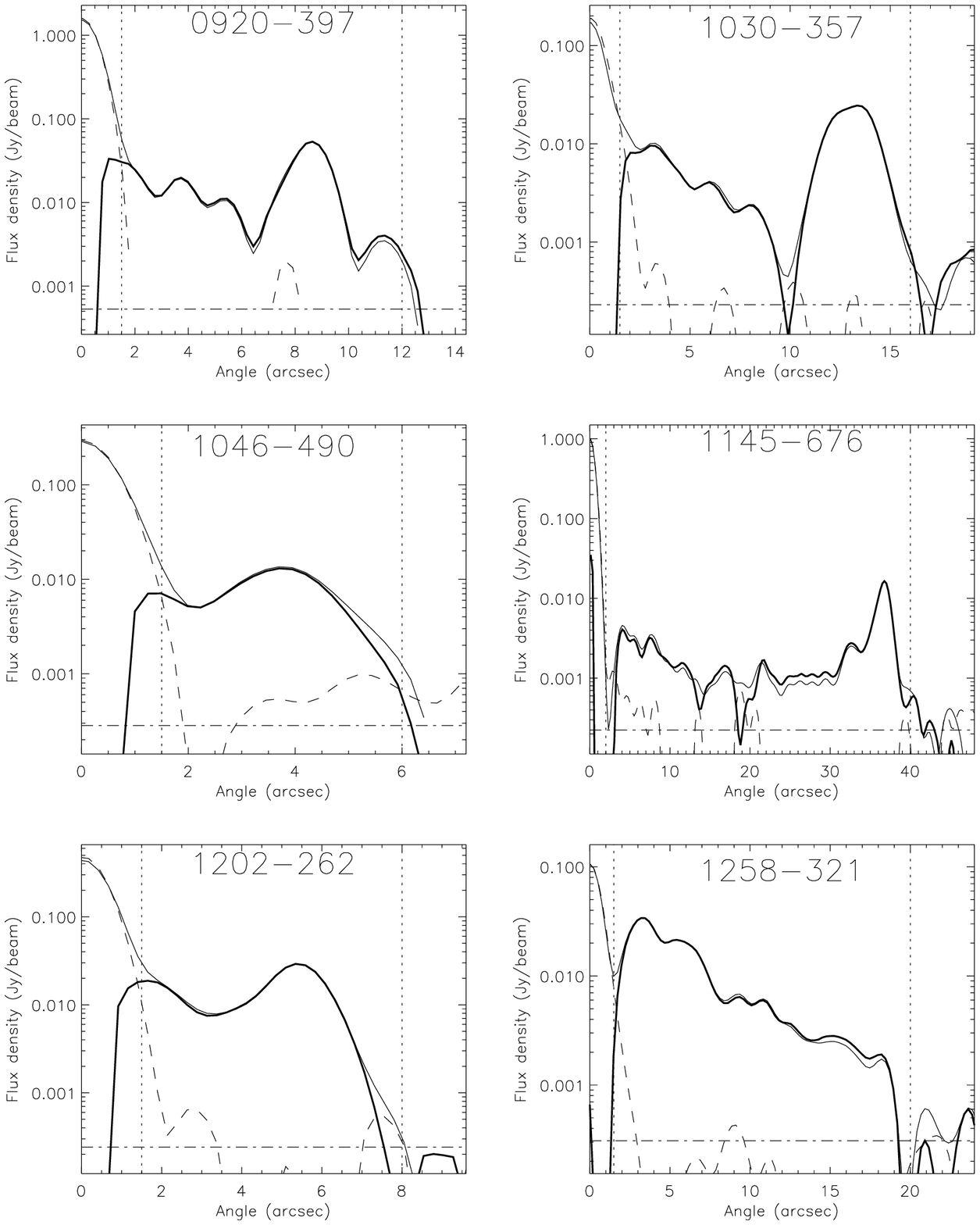}
  \caption{continued.}
\end{figure}

\addtocounter{figure}{-1}

\begin{figure}[htp]
\plotone{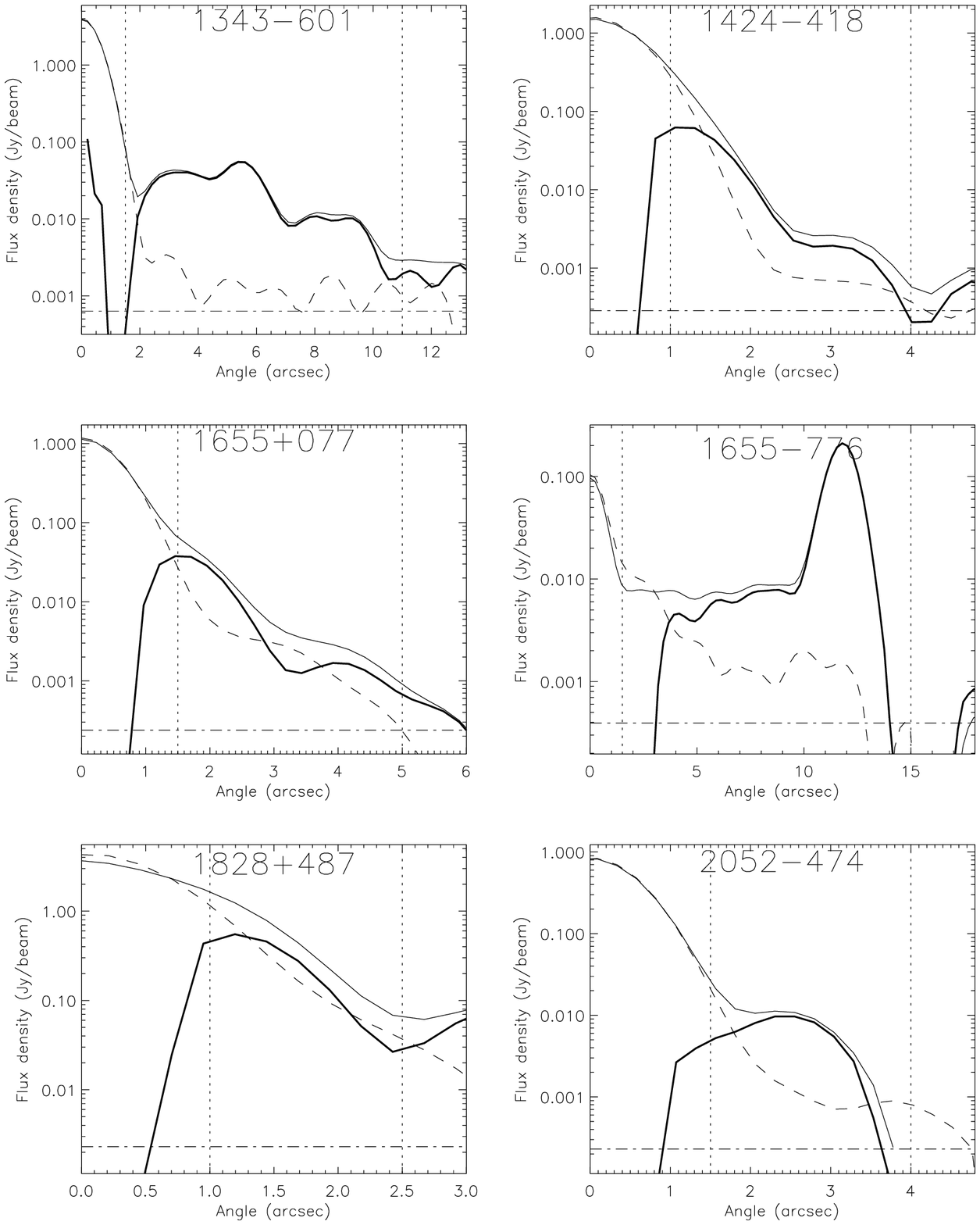}
  \caption{continued.}
\end{figure}

\addtocounter{figure}{-1}

\begin{figure}[htp]
\plotone{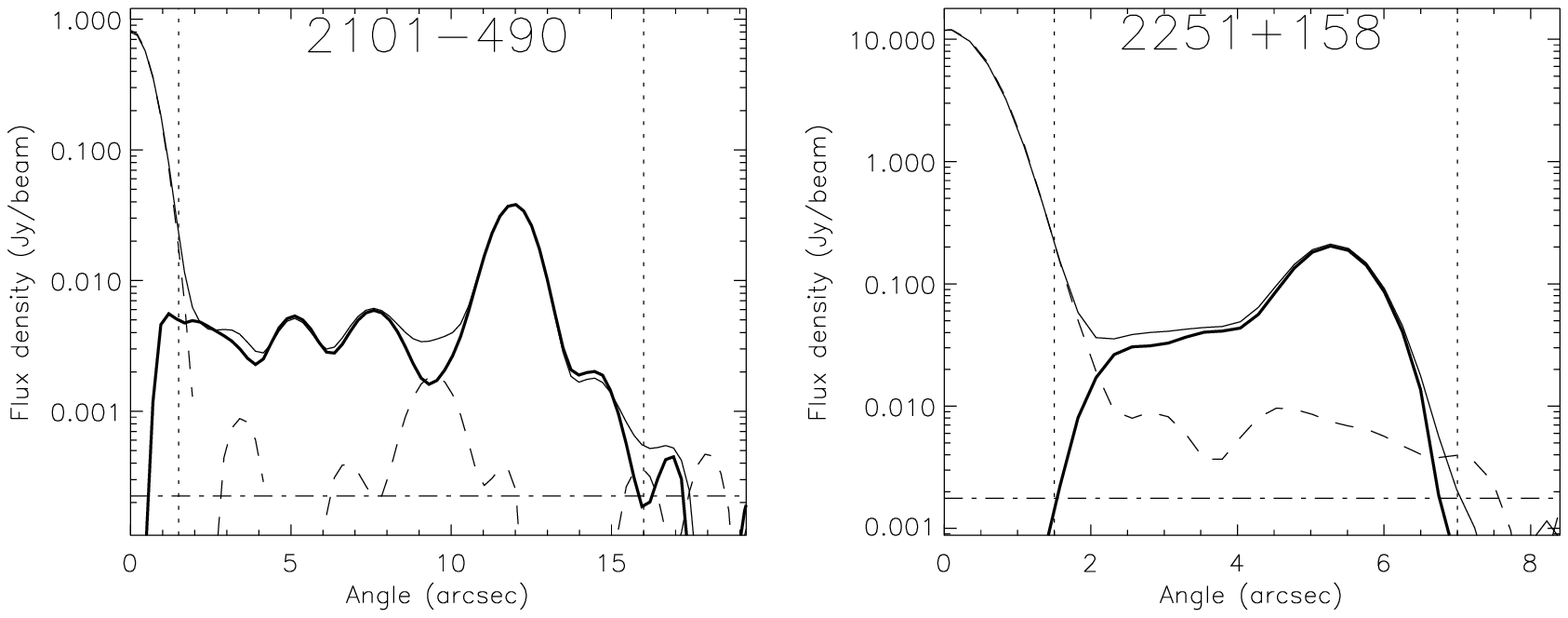}
  \caption{continued.}
\end{figure}

\begin{figure}[htp]
\plotone{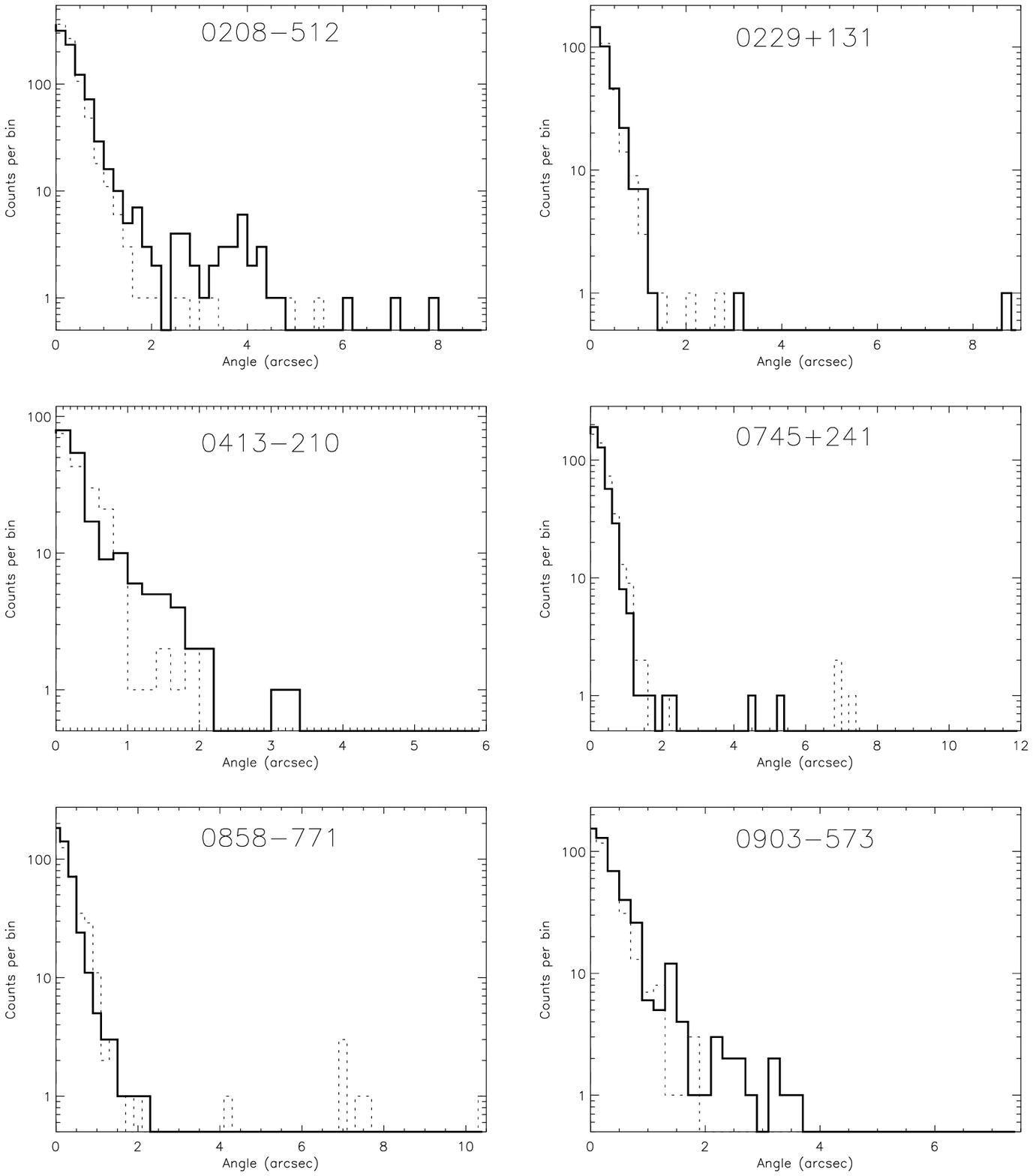}
  \caption{Histograms of counts from the X-ray images in
  0.2\arcsec\ bins.  The solid, bold
  lines give the profiles along the position angles of the jets, as defined
  in Table~\ref{tab:jetresults} and used in Fig.~\ref{fig:radioprofiles}.
  The dashed histograms give the profiles at
  a position angle 180\arcdeg\ opposite to the jet -- the counter-jet
  direction.  The counter-jet profiles provide a measure of the significance
  of the X-ray emission from the jet because there are no clearly detected
  counter-jets.} \label{fig:xrayprofiles}
\end{figure}

\addtocounter{figure}{-1}

\begin{figure}[htp]
\plotone{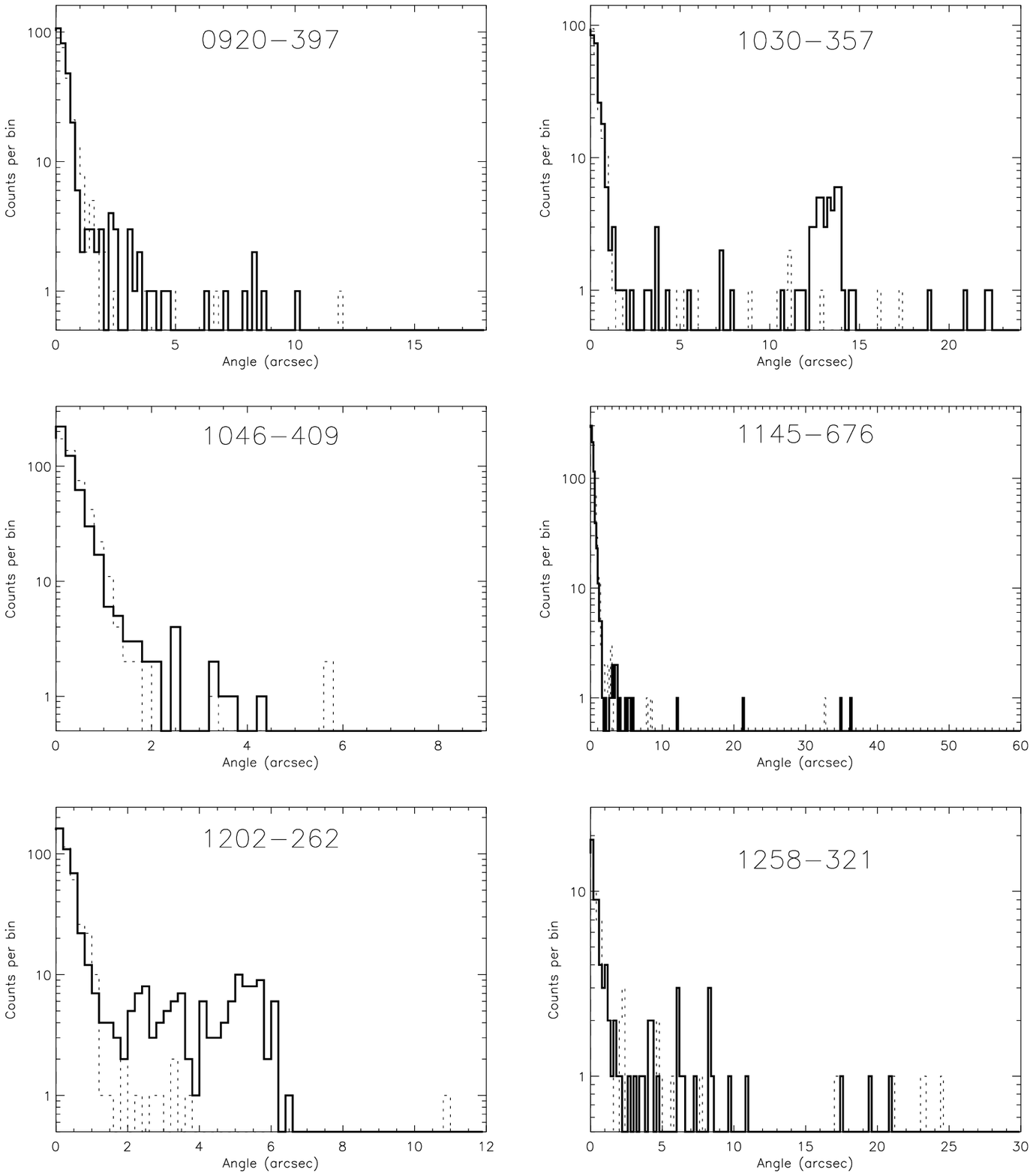}
  \caption{continued.}
\end{figure}

\addtocounter{figure}{-1}

\begin{figure}[htp]
\plotone{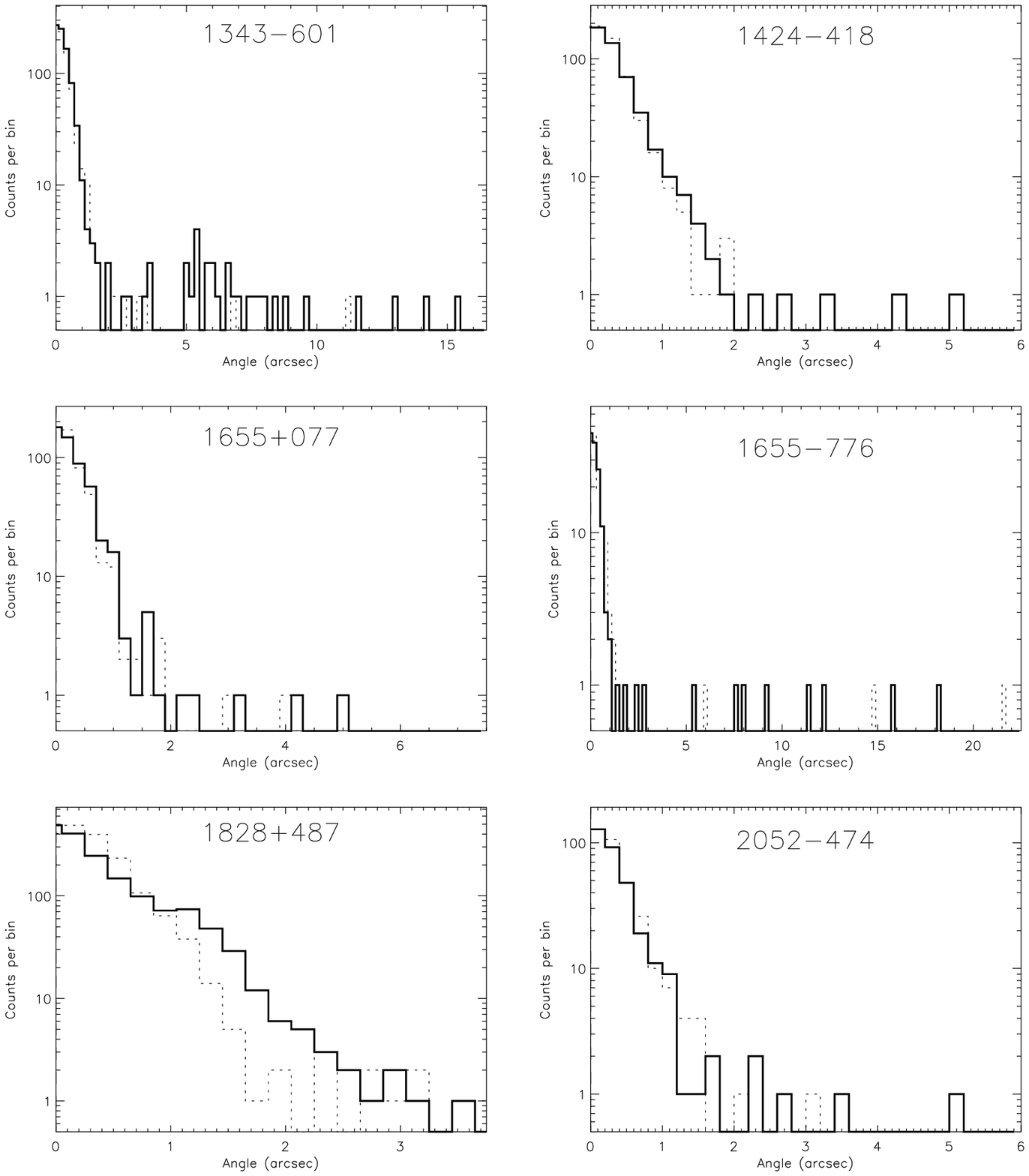}
  \caption{continued.}
\end{figure}

\addtocounter{figure}{-1}

\begin{figure}[htp]
\plotone{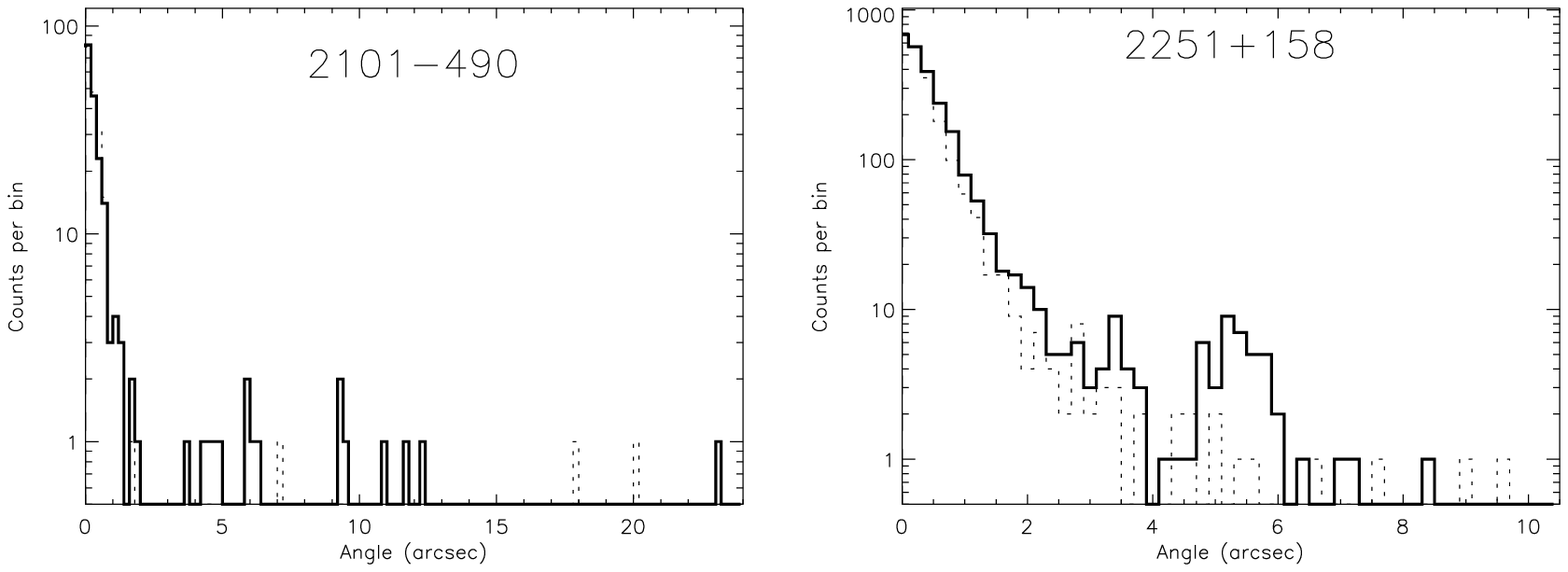}
  \caption{continued.}
\end{figure}

\begin{figure}
\begin{center}
\plotone{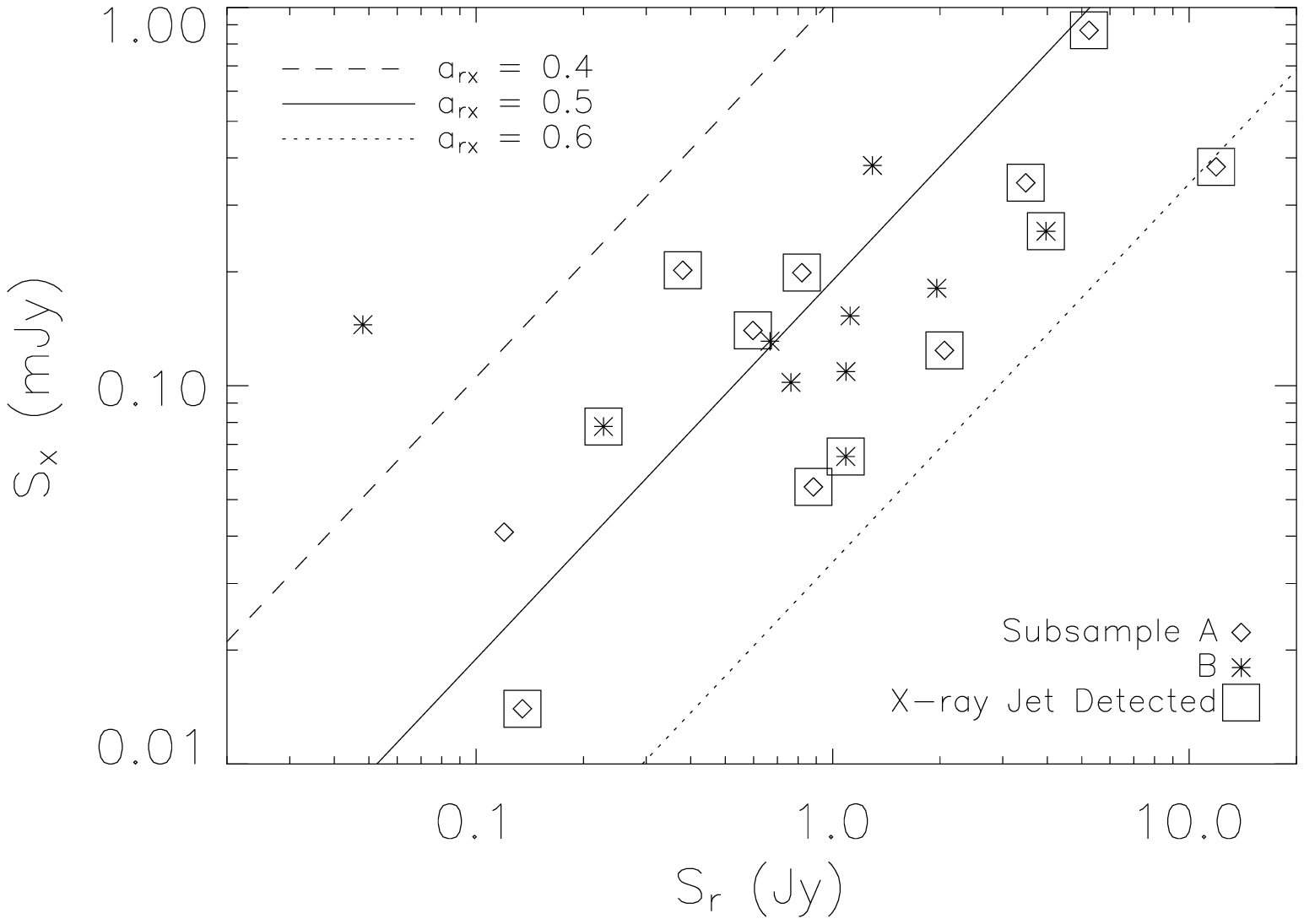}
\caption{X-ray and radio flux densities for the cores of the targets
in the observation set.  The radio flux densities were measured from
the maps shown in Fig.~\ref{fig:images} and then adjusted to 5 GHz
assuming $\alpha_{\rm r} = 0.5$.
The X-ray flux densities are estimated at 1 keV ($2.42 \times 10^{17}$ Hz).
Error bars in either flux density are generally
small -- $\lae$ 10\%.  The A and B targets have similar distributions.
The spectral indices from the radio to the X-ray band ($\alpha_{rx}$)
cluster about a value of 0.5.  Similarly, jets are detected without
bias regarding either this spectral index
or the radio or X-ray core flux densities.
\label{fig:corefluxes} }
\end{center}
\end{figure}

\begin{figure}
\begin{center}
\plotone{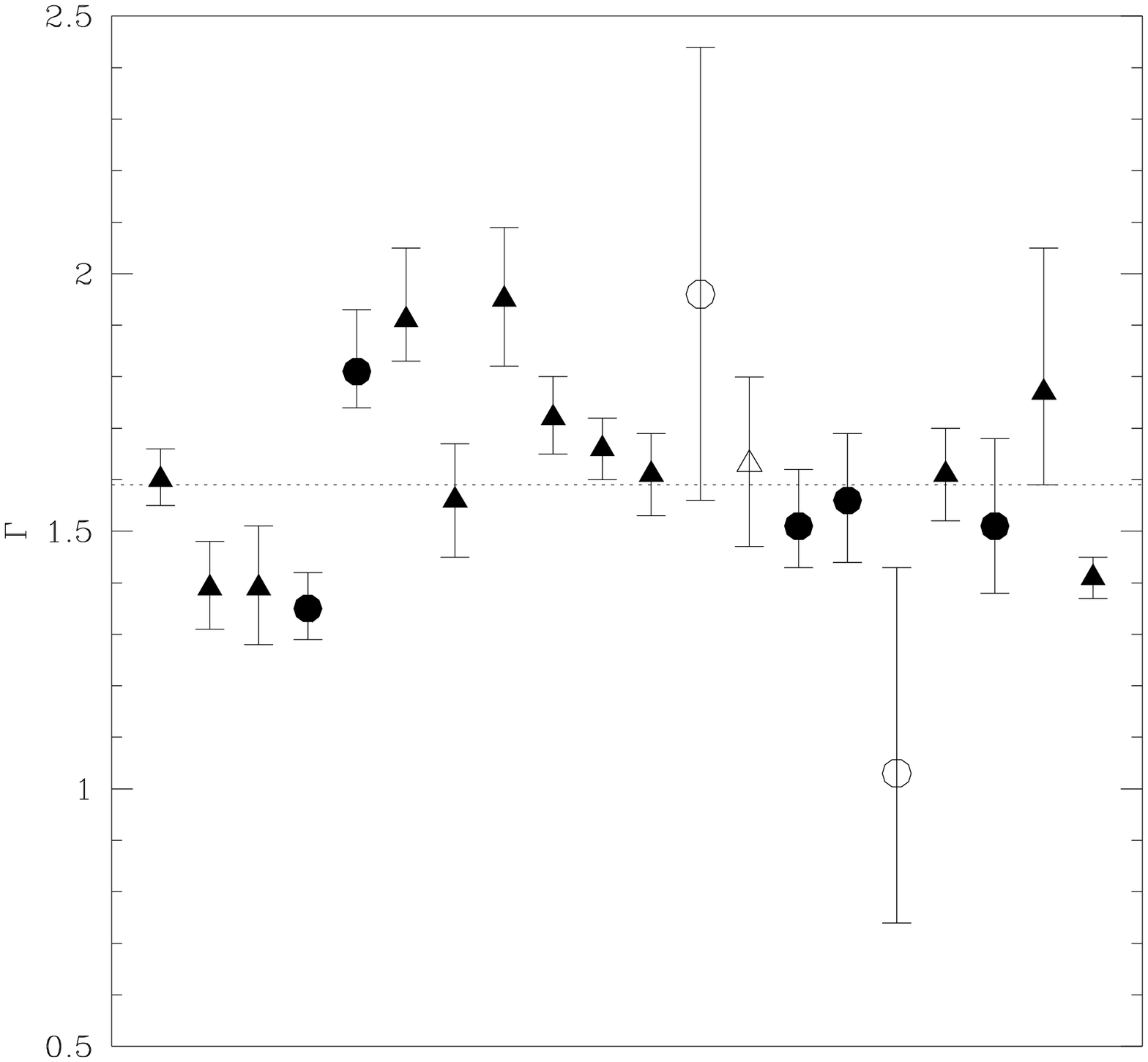}
\caption{X-ray spectral indices of the cores.  Objects are in RA
order, as in Table~\ref{tab:coreresults}.  Objects with X-ray jet emission are
shown as triangles, and those without as circles.  Unfilled symbols
mark the three low-redshift objects of our sample.
The dotted horizontal line is the best-fit photon index
(given by $\Gamma = \alpha_{\rm x} + 1$).
\label{fig:xrayspectra} }
\end{center}
\end{figure}

\begin{figure}
\begin{center}
\plotone{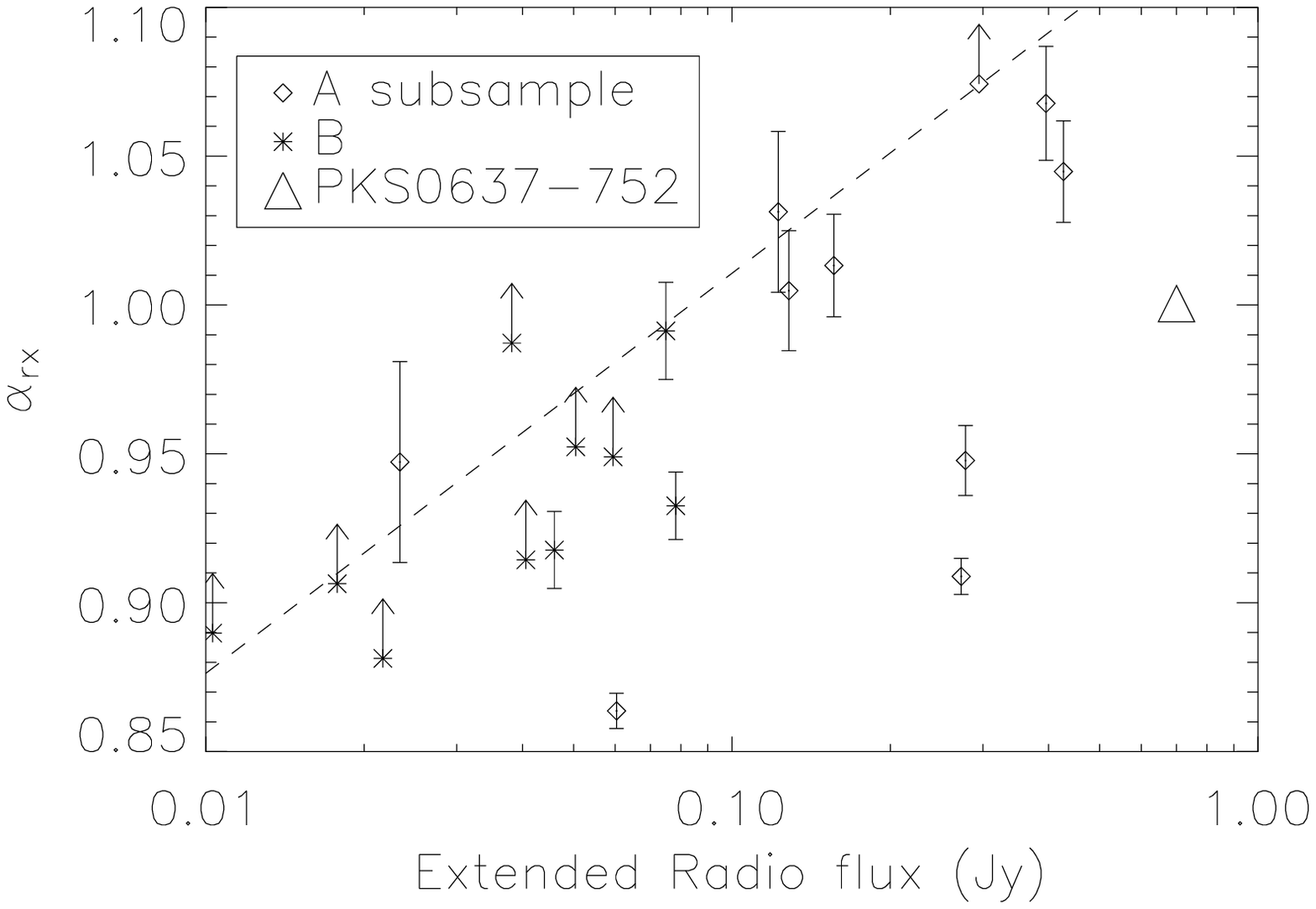}
\caption{Plot of $\alpha_{rx}$ against radio flux density in
the extended emission.  The result for \pks\ is given for
comparison.
The dashed line gives the limit to $\alpha_{rx}$ for a hypothetical
observation where the limit to the X-ray count rate is 0.003 count/s.
Actual limits vary from source to source due to the varying angular
sizes and locations of the radio jets.  Sources included in our survey
that satisfy only the subsample B category (based on morphology) are
more likely to be X-ray faint, which reduces the limiting value of $\alpha_{rx}$.
\label{fig:alpharx} }
\end{center}
\end{figure}

\begin{figure}
\begin{center}
\plotone{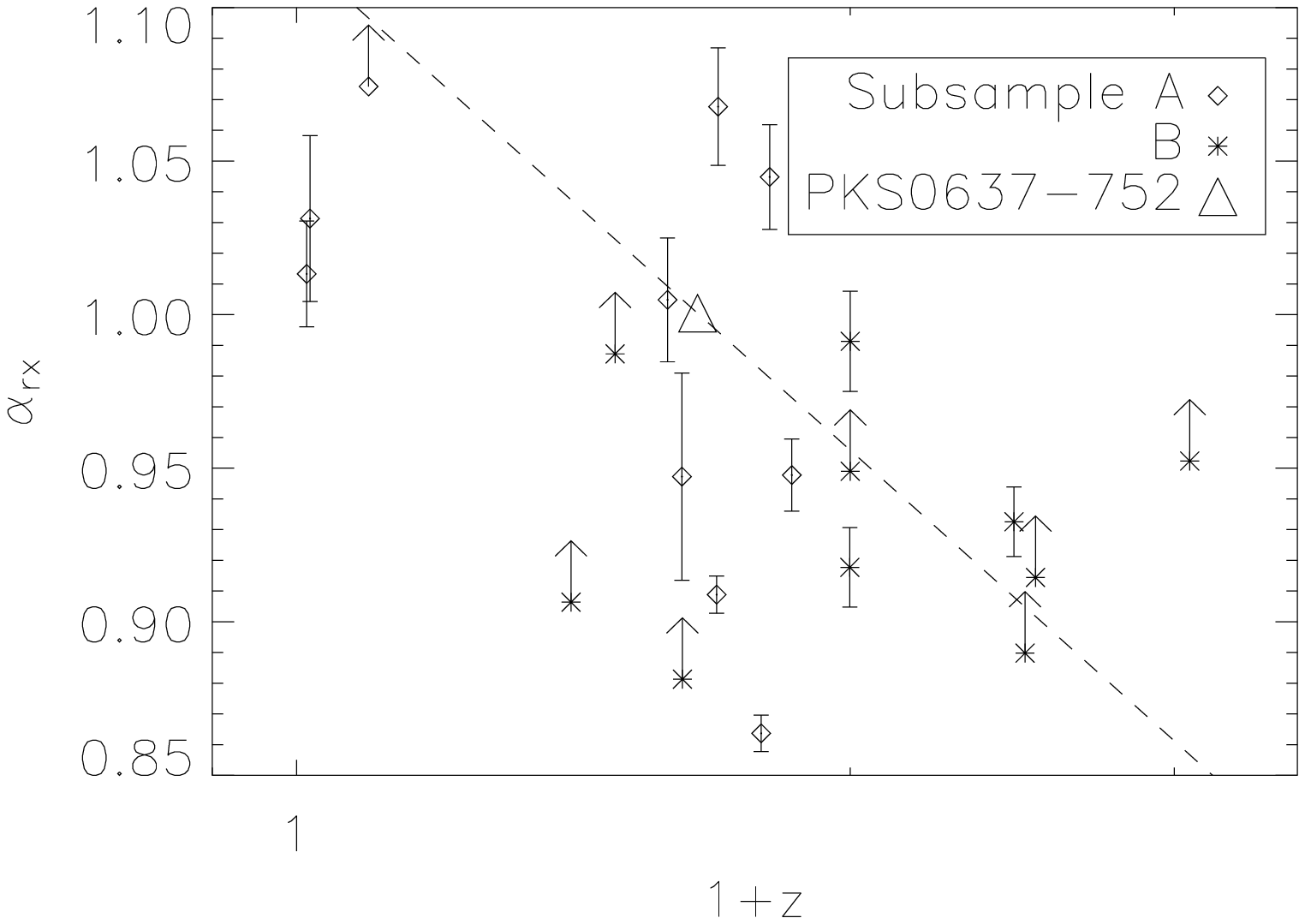}
\caption{Plot of $\alpha_{rx}$ against redshift.
The result for \pks\ is given for comparison.
The dashed line gives the dependence of $\alpha_{rx}$ on $z$ under
the assumptions that the X-ray emission results only from inverse Compton
scattering off of the cosmic microwave background and that the beaming parameters
for all jets are the same as those of \pks, so that the X-ray
flux density would increase as $(1+z)^4$.  Clearly, there is a wide distribution
of the observed values of $\alpha_{rx}$, indicating that the beaming parameters
vary widely.  We set $z=1$ for the source without a measured redshift.
\label{fig:alpharx-z} }
\end{center}
\end{figure}

\clearpage

\begin{deluxetable}{lllllrcc}
\tablewidth{0pc}
\tabletypesize{\scriptsize}
\tablecaption{Quasars in the Complete Sample \label{tab:sample} }
\tablehead{
\colhead{B1950} & \colhead{Name} & \colhead{$z$\tablenotemark{a}} & \colhead{$m_V$}
	& \multicolumn{2}{c}{$S_5$} & \multicolumn{2}{c}{Subsample} \\
\colhead{} & \colhead{} & \colhead{} & \colhead{} & \colhead{Core}
	& \colhead{Jet\tablenotemark{b}} & \colhead{(A)} & \colhead{(B)} \\
\colhead{} & \colhead{} & \colhead{} & \colhead{} & \colhead{(Jy)}
	& \colhead{(mJy)} & \colhead{} & \colhead{} }
\startdata
0106$+$013	&	4C +01.02	&	2.099	&	18.5	&	3.841	&	198	&	A	&	B	\\
0144$-$522	&	PKS	&	-	&	17.5	&	0.189	&	47	&	-	&	B	\\
0208$-$512	&	PKS	&	0.999	&	17	&	3.053	&	127	&	-	&	B	\\
0229$+$131	&	4C +13.14	&	2.059	&	17.5	&	1.118	&	100	&	-	&	B	\\
0234$+$285	&	4C +28.04	&	1.213	&	19	&	2.464	&	50	&	-	&	B	\\
0256$+$075	&	PKS	&	0.893	&	18.5	&	0.552	&	18	&	-	&	B	\\
0402$-$362	&	PKS	&	1.417	&	17	&	1.838	&	19	&	-	&	B	\\
0413$-$210	&	PKS	&	0.808	&	19	&	1.027	&	342	&	A	&	B	\\
0454$-$463	&	PKS	&	0.858	&	18	&	1.088	&	289	&	A	&	-	\\
0508$-$220	&	PKS	&	0.1715	&	18.5	&	0.107	&	249	&	A	&	-	\\
0707$+$476	&		&	1.292	&	18	&	0.973	&	44	&	-	&	B	\\
0745$+$241	&		&	0.410	&	19	&	0.719	&	92	&	-	&	B	\\
0748$+$126	&		&	0.889	&	18	&	1.43	&	13	&	-	&	B	\\
0820$+$225	&	4C +22.21	&	0.951	&	19.5	&	1.606	&	283	&	A	&	B	\\
0833$+$585	&		&	2.101	&	18	&	0.678	&	14	&	-	&	B	\\
0858$-$771	&	PKS	&	0.490	&	17.5	&	0.192	&	117	&	-	&	B	\\
0859$+$470	&	4C +47.29	&	1.462	&	19	&	1.645	&	149	&	A	&	B	\\
0903$-$573	&	PKS	&	0.695	&	18.5	&	0.746	&	563	&	A	&	B	\\
0920$-$397	&	PKS	&	0.591	&	18.5	&	1.29	&	303	&	A	&	-	\\
0923$+$392	&	4C +39.25	&	0.6948	&	18	&	2.681	&	308	&	A	&	-	\\
0953$+$254	&		&	0.712	&	17	&	0.483	&	9	&	-	&	B	\\
0954$+$556	&	4C +55.17	&	0.909	&	17.5	&	2.568	&	179	&	A	&	B	\\
1030$-$357	&	PKS	&	1.455	&	19	&	0.27	&	121	&	-	&	B	\\
1040$+$123	&	3C 245	&	1.029	&	17	&	1.627	&	554	&	A	&	-	\\
1046$-$409	&	PKS	&	0.620	&	18	&	0.494	&	404	&	A	&	B	\\
1055$+$018	&	4C +01.28	&	0.888	&	18.5	&	2.642	&	76	&	-	&	B	\\
1055$+$201	&	4C +20.24	&	1.110	&	17	&	0.768	&	858	&	A	&	B	\\
1116$+$128	&	4C +12.39	&	2.118	&	19	&	1.888	&	122	&	A	&	B	\\
1116$-$462	&	PKS	&	0.713	&	17	&	1.234	&	289	&	A	&	B	\\
1145$-$676	&	PKS	&	-	&	18	&	1.238	&	137	&	-	&	B	\\
1202$-$262	&	PKS	&	0.789	&	19.5	&	0.593	&	334	&	A	&	B	\\
1251$-$713	&	PKS	&	-	&	21	&	1.432	&	75	&	-	&	B	\\
1258$-$321	&	ESO 443-G024; MCG -5-31-12	&	0.01704	&	13	&	0.141	&	287	&	A	&	-	\\
1303$-$827	&	PKS	&	-	&	17.5	&	0.221	&	260	&	A	&	-	\\
1343$-$601	&	PKS, Cen B	&	0.01292	&	11	&	2.34	&	585	&	A	&	B	\\
1354$+$195	&	4C +19.44	&	0.720	&	16	&	1.309	&	402	&	A	&	B	\\
1421$-$490	&	PKS	&	-	&	24	&	5.261	&	277	&	A	&	B	\\
1424$-$418	&	PKS	&	1.522	&	18	&	2.138	&	113	&	-	&	B	\\
1502$+$106	&		&	1.839	&	18.5	&	2.198	&	11	&	-	&	B	\\
1622$-$297	&	PKS	&	0.815	&	19.5	&	2.041	&	42	&	-	&	B	\\
1641$+$399	&	3C 345; 4C +39.48	&	0.5928	&	16	&	8.5	&	475	&	A	&	B	\\
1642$+$690	&	4C +69.21	&	0.751	&	19	&	0.999	&	155	&	A	&	B	\\
1655$+$077	&		&	0.621	&	20.5	&	1.165	&	99	&	-	&	B	\\
1655$-$776	&	PKS	&	0.0944	&	17.5	&	0.425	&	563	&	A	&	-	\\
1823$+$568	&	4C +56.27	&	0.664	&	19	&	0.858	&	212	&	A	&	B	\\
1828$+$487	&	3C 380; 4C +48.46	&	0.692	&	17	&	4.5	&	3657	&	A	&	-	\\
1928$+$738	&	1ES, 4C +73.18	&	0.3021	&	16.5	&	2.939	&	114	&	A	&	B	\\
2007$+$777	&		&	0.342	&	16.5	&	0.823	&	14	&	-	&	B	\\
2052$-$474	&	PKS	&	1.489	&	18.5	&	1.103	&	58	&	-	&	B	\\
2101$-$490	&	PKS	&	(1.04)	&	17	&	0.524	&	184	&	-	&	B	\\
2123$-$463	&	PKS	&	1.67	&	18.5	&	0.546	&	75	&	-	&	B	\\
2201$+$315	&	4C +31.63	&	0.295	&	15.5	&	1.4	&	180	&	A	&	B	\\
2230$+$114	&	4C +11.69	&	1.037	&	17.5	&	6.468	&	134	&	A	&	B	\\
2251$+$158	&	3C 454.3; 4C +15.76	&	0.859	&	16	&	12.19	&	355	&	A	&	B	\\
2255$-$282	&	PKS	&	0.926	&	17	&	1.602	&	67	&	-	&	B	\\
2326$-$477	&	PKS	&	1.299	&	17	&	1.679	&	52	&	-	&	B	\\
\enddata
\tablenotetext{a}{Redshifts were obtained from the NASA Extragalactic Database except
for 2101$-$490, which is taken from \citet{gm04}.}
\tablenotetext{b}{For the jets, $S_{\rm r}$ is the computed radio flux density of the
extended emission based on
the ATCA and VLA images and converted to 5 GHz.}
\end{deluxetable}

\begin{deluxetable}{lccc}
\tablecolumns{4}
\tablewidth{0pc}
\tabletypesize{\scriptsize}
\tablecaption{Observation Log \label{tab:observations} }
\tablehead{
\colhead{Target} & \colhead{{\em Chandra}} & \colhead{Live Time} 
	& \colhead{Date} \\
\colhead{} & \colhead{Obs ID} & \colhead{(s)} 
	& \colhead{(UT)} }
\startdata
0208$-$512 &  3108 & 5014 & 2001-12-13 \\
0229$+$131 &  3109 & 5410 & 2002-01-30 \\
0413$-$210 &  3110 & 4862 & 2002-04-08 \\
0745$+$241 &  3111 & 5019 & 2002-03-07 \\
0858$-$771 &  3112 & 4962 & 2002-06-19 \\
0903$-$573 &  3113 & 4934 & 2002-07-04 \\
0920$-$397 &  3114 & 4466 & 2002-07-04 \\
1030$-$357 &  3115 & 5539 & 2002-05-07 \\
1046$-$409 &  3116 & 4330 & 2002-04-24 \\
1145$-$676 &  3117 & 4621 & 2002-04-24 \\
1202$-$262 &  3118 & 5074 & 2002-03-02 \\
1258$-$321 &  3119 & 5426 & 2002-03-08 \\
1343$-$601 &  3120 & 5160 & 2002-07-24 \\
1424$-$418 &  3121 & 4475 & 2002-07-04 \\
1655$+$077 &  3122 & 4825 & 2002-04-28 \\
1655$-$776 &  3123 & 4917 & 2002-05-21 \\
1828$+$487 &  3124 & 5329 & 2002-05-20 \\
2052$-$474 &  3125 & 5310 & 2002-07-29 \\
2101$-$490 &  3126 & 5370 & 2002-08-02 \\
2251$+$158 &  3127 & 4929 & 2002-11-06 \\
\enddata
\end{deluxetable}

\begin{deluxetable}{lcclc}
\tablecolumns{5}
\tablewidth{0pc}
\tabletypesize{\scriptsize}
\tablecaption{Radio contours \label{tab:radioCont} }
\tablehead{
\colhead{Target} & \colhead{Instrument} & \colhead{Date}
	& \colhead{Freq.} & \colhead{$5 \times$ RMS noise} \\
\colhead{} & \colhead{} & \colhead{(UT)} 
	& \colhead{(GHz)} & \colhead{(mJy/beam)} }
\startdata
0208$-$512 & \mbox{\em ATCA} & 2002-02-01 & 8.64  & 1.275  \\
0229$+$131 & \mbox{\em  VLA} & 2000-11-05 & 1.425 & 1.065  \\
0413$-$210 & \mbox{\em  VLA} & 2000-11-05 & 4.86  & 2.540  \\
0745$+$241 & \mbox{\em  VLA} & 2001-05-06 & 4.86  & 0.695  \\
0858$-$771 & \mbox{\em ATCA} & 2000-05-20 & 8.64  & 0.495  \\
0903$-$573 & \mbox{\em ATCA} & 2002-02-04 & 8.64  & 1.345  \\
0920$-$397 & \mbox{\em ATCA} & 2000-09-01 & 8.64  & 2.300  \\
1030$-$357 & \mbox{\em ATCA} & 2000-05-22 & 8.64  & 0.390  \\
1030$-$357\tablenotemark{a} & \mbox{\em ATCA} & 2000-05-22 & 4.80  & 0.645  \\
1046$-$409 & \mbox{\em ATCA} & 2002-05-22 & 8.64  & 0.820  \\
1145$-$676 & \mbox{\em ATCA} & 2002-02-02 & 8.64  & 0.670  \\
1202$-$262 & \mbox{\em ATCA} & 2000-05-22 & 8.64  & 0.500  \\
1258$-$321 & \mbox{\em ATCA} & 2002-02-02 & 8.64  & 0.605  \\
1343$-$601\tablenotemark{b} & \mbox{\em ATCA} & 2002-02-02 & 8.64  & 2.155  \\
1424$-$418 & \mbox{\em ATCA} & 2002-02-04 & 8.64  & 1.250  \\
1655$+$077 & \mbox{\em  VLA} & 2001-05-06 & 1.46  & 0.835  \\
1655$-$776 & \mbox{\em ATCA} & 2002-02-05 & 8.64  & 0.830  \\
1828$+$487 & \mbox{\em  VLA} & 2000-11-05 & 4.86  & 1.079  \\
2052$-$474 & \mbox{\em ATCA} & 2002-02-05 & 8.64  & 0.825  \\
2101$-$490 & \mbox{\em ATCA} & 2000-05-25 & 8.64  & 0.610  \\
2251$+$158 & \mbox{\em  VLA} & 2001-05-06 & 4.86  & 8.100  \\
\enddata
\tablenotetext{a}{For 1030$-$357 (Fig.~\ref{fig:images}h), 
  4.8~GHz radio contours at $5\times$ and $10\times$
  the RMS level are overlaid in red to show large-scale structure.}
\tablenotetext{b}{Fig.~\ref{fig:images}m includes an extra contour at
  1.293 mJy/beam ($3\times$ the RMS level).}
\end{deluxetable}

\begin{deluxetable}{lccccccc}
\tablecolumns{8}
\tablewidth{0pc}
\tabletypesize{\scriptsize}
\tablecaption{Quasar X-ray Core Measurements \label{tab:coreresults} }
\tablehead{
\colhead{B1950} & \colhead{Count Rate}
	& \colhead{Streak Rate} & \colhead{$N_{\rm H_{\rm Gal}}$\tablenotemark{a}}  & \colhead{$\Gamma_{\rm x}$} 
	& \colhead{$N_{\rm H_{\rm int}}$} 
	& \colhead{$S_{\rm x}$\tablenotemark{b}} & \colhead{$\chi^2/$(dof)} \\
\colhead{} & \colhead{(cps)} & \colhead{(cps)} & \colhead{($10^{21}$ cm$^{-2}$)} & \colhead{} 
	& \colhead{($10^{22}$ cm$^{-2}$)} & \colhead{(nJy)} & \colhead{} }
\startdata
0208$-$512&0.305 $\pm$ 0.008& 0.54 $\pm$ 0.15& 0.294 & $1.60^{+0.06}_{-0.05}$ & $< 0.34$        & $256^{+15}_{-8}$& 56.2/41 \\ 
0229$+$131&0.111 $\pm$ 0.005& 0.08 $\pm$ 0.10& 0.83  & $1.39^{+0.09}_{-0.08}$ & $< 2.13$        & 102 $\pm$  7  & 16.7/16 \\ 
0413$-$210&0.063 $\pm$ 0.004& 0.09 $\pm$ 0.11& 0.239 & $1.39^{+0.12}_{-0.11}$ & $< 0.69$        & $54^{+5}_{-4}$ & 4.1/7 \\ 
0745$+$241&0.160 $\pm$ 0.006& 0.23 $\pm$ 0.16& 0.516 & $1.35^{+0.07}_{-0.06}$ & $< 0.19$        & $131^{+7}_{-6}$  & 23.5/22 \\ 
0858$-$771&0.130 $\pm$ 0.005& 0.52 $\pm$ 0.21& 1.021 & $1.81^{+0.12}_{-0.07}$ & $< 0.37$        & $145^{+15}_{-7}$ & 18.1/17 \\ 
0903$-$573&0.123 $\pm$ 0.005& 0.00 $\pm$ 0.08& 3.212 & $1.91^{+0.14}_{-0.08}$ & $< 0.74$        & $199^{+26}_{-12}$ & 14.9/16 \\ 
0920$-$397&0.105 $\pm$ 0.005& 0.11 $\pm$ 0.12& 2.147 & 1.56 $\pm$ 0.11        & $< 0.48$        & $124^{+10}_{-9}$   & 23.3/12 \\ 
1030$-$357&0.069 $\pm$ 0.004& 0.14 $\pm$ 0.14& 0.609 & $1.95^{+0.14}_{-0.13}$ & $< 1.48$        & $78^{+7}_{-5}$ & 6.0/8 \\ 
1046$-$409&0.196 $\pm$ 0.007&-0.01 $\pm$ 0.09& 0.829 & $1.72^{+0.08}_{-0.07}$ & $< 0.25$        & $202^{+11}_{-9}$ & 16.4/23 \\ 
1145$-$676\tablenotemark{c}&0.292 $\pm$ 0.008& 0.49 $\pm$ 0.23& 3.189 & 1.66 $\pm$ 0.06        & $< 0.08$        & $382^{+17}_{-16}$ & 69.2/38 \\ 
1202$-$262\tablenotemark{d}&0.147 $\pm$ 0.005& 0.16 $\pm$ 0.13& 0.708 & 1.61 $\pm$ 0.08        & $< 0.45$        & $140^{+10}_{-8}$ & 21.8/20 \\ 
1258$-$321\tablenotemark{e}&0.011 $\pm$ 0.002& 0.09 $\pm$ 0.16& 0.575 & $1.96^{+0.48}_{-0.40}$ & $< 1.76$        & $14^{+6}_{-3}$ & 6.5/2 \\ 
1343$-$601\tablenotemark{e}&0.252 $\pm$ 0.007& 0.53 $\pm$ 0.27& 10.6  & $1.63^{+0.17}_{-0.16}$ & 1.08 $\pm$ 0.24 & $870^{+240}_{-180}$ &35.1/35 \\
1424$-$418&0.189 $\pm$ 0.007& 0.08 $\pm$ 0.12& 0.805 & $1.51^{+0.11}_{-0.08}$ & $< 1.37$        & $181^{+20}_{-11}$ & 13.8/23 \\ 
1655$+$077&0.165 $\pm$ 0.006& 0.09 $\pm$ 0.14& 0.626 & $1.56^{+0.13}_{-0.12}$ & $< 0.55$        & $153^{+21}_{-16}$ & 38.3/22 \\ 
1655$-$776\tablenotemark{e}&0.031 $\pm$ 0.003& 0.02 $\pm$ 0.11& 0.830 & $1.03^{+0.40}_{-0.29}$ & $< 3.17$        & $41^{+23}_{-9}$ & 4.2/2 \\ 
1828$+$487\tablenotemark{f}&0.421 $\pm$ 0.009& 0.73 $\pm$ 0.25& 0.66 & 1.61 $\pm$ 0.09 & $< 0.09$ & 344 $\pm$  10 & 74.6/61 \\ 
2052$-$474&0.113 $\pm$ 0.005& 0.48 $\pm$ 0.21& 0.404 & $1.51^{+0.17}_{-0.13}$ & $< 1.96$        & $109^{+18}_{-10}$ & 10.3/15 \\ 
2101$-$490&0.060 $\pm$ 0.003& 0.07 $\pm$ 0.11& 0.341 & $1.77^{+0.28}_{-0.18}$ & $< 0.28$        & $65^{+17}_{-9}$ & 1.2/7 \\ 
2251$+$158\tablenotemark{g}&0.656 $\pm$ 0.012& 2.89 $\pm$ 0.50& 0.713 & 1.41 $\pm$ 0.04 & $< 0.12$ & 379 $\pm$ 11  & 80.5/89 \\ 
\enddata
\tablenotetext{a}{From the COLDEN program provided by the CXC, using
data from Dickey \& Lockman (1990), except for 0229$+$131 (from Murphy et al. 1996) and
1828$+$487 and 2251$+$158 (from Elvis et al. 1989)}
\tablenotetext{b}{$S_{\rm x}$ is the flux density at 1 keV from spectral fits.
One may roughly estimate $S_{\rm x}$ by scaling the count rate by 1000
nJy/(count/s).}
\tablenotetext{c}{Indication of a soft excess, not modelled here.}
\tablenotetext{d}{The inclusion of a narrow K$\alpha$ fluorescence line from
neutral Fe at a (fixed) rest energy of 6.4 keV improves the fit
significantly (5\% probability
of the improvement in $\chi^2$ being by chance, on an F-test).  The equivalent width is
$221^{+115}_{-109}$ eV.}
\tablenotetext{e}{Low-redshift object not included in statistical
analysis of the spectral results.}
\tablenotetext{f}{Spectrum shows signs of pileup. Quoted results for $\chi^2$ and
the power-law slope are with the
pile-up model in XSPEC applied.  Without it the
fit is unacceptable: $\chi^2 = 84$, $\Gamma = 1.39 \pm 0.04$.}
\tablenotetext{g}{Spectrum shows signs of pileup. Quoted results for $\chi^2$ and
the power-law slope are with the
pile-up model in XSPEC applied.  Without it the
fit is unacceptable: $\chi^2 = 116$, $\Gamma = 0.79 \pm 0.03$.}
\tablecomments{Results correspond to the energy band 0.5-7 keV.  Upper limits are $3\sigma$ confidence.  
Otherwise uncertainties are $1\sigma$ for one interesting parameter.
Data have been grouped to a minimum of 30 counts per bin before
fitting, except for 1258$-$321 for which statistics are poor, and the
grouping is a minimum of 15 counts per bin.
}
\end{deluxetable}

\begin{deluxetable}{rrrrrrrrrrr}
\tablecolumns{11}
\tablewidth{0pc}
\tabletypesize{\scriptsize}
\tablecaption{Quasar Jet Measurements \label{tab:jetresults} }
\tablehead{
\colhead{Target} & \colhead{PA} 
	& \colhead{$\theta_i$} & \colhead{$\theta_o$} 
	& \colhead{$S_{\rm r}$\tablenotemark{a}} & \colhead{$\nu_{\rm r}$} & \colhead{Count Rate} & \colhead{$S_{\rm x}$\tablenotemark{a}} 
	& \colhead{$\alpha_{rx}$} & \colhead{$P_{jet}$\tablenotemark{b}} & \colhead{X?\tablenotemark{c}} \\
\colhead{ } & \colhead{(\arcdeg)} 
	& \colhead{(\arcsec)} & \colhead{(\arcsec)} 
	& \colhead{(mJy)} & \colhead{(GHz)} & \colhead{(cps)} & \colhead{(nJy)} & \colhead{} 
	& \colhead{} & \colhead{} }
\startdata
0208$-$512 & -130 &  1.5 &  6.0 &  46.0 $\pm$ 1.0 & 8.64 &  0.0068 $\pm$ 0.0015 &      6.8 &     0.92 $\pm$ 0.01 & 1.33e-14 & Y \\
0229$+$131 &   20 &  1.5 &  6.0 &  50.5 $\pm$ 0.7 & 1.42 & -0.0004 $\pm$ 0.0004 & $<$  0.7 & $>$ 0.95            & 9.50e-01 & N \\
0413$-$210 &  155 &  1.0 &  4.0 & 426.9 $\pm$ 1.2 & 4.86 &  0.0039 $\pm$ 0.0012 &      3.9 &     1.04 $\pm$ 0.02 & 2.85e-08 & Y \\
0745$+$241 &  -45 &  1.5 &  8.0 &  17.8 $\pm$ 0.8 & 4.86 &  0.0008 $\pm$ 0.0007 & $<$  2.9 & $>$ 0.88            & 5.11e-02 & N \\
0858$-$771 &  165 &  1.5 &  7.0 &  38.2 $\pm$ 0.6 & 8.64 &  0.0000 $\pm$ 0.0006 & $<$  1.7 & $>$ 0.99            & 5.67e-01 & N \\
0903$-$573 &   35 &  1.0 &  5.0 & 395.4 $\pm$ 0.8 & 8.64 &  0.0045 $\pm$ 0.0015 &      4.5 &     1.07 $\pm$ 0.02 & 1.19e-06 & Y \\
0920$-$397 &  180 &  1.5 & 12.0 & 128.3 $\pm$ 2.3 & 8.64 &  0.0043 $\pm$ 0.0015 &      4.3 &     1.00 $\pm$ 0.02 & 3.37e-06 & Y \\
1030$-$357 & -150 &  1.5 & 16.0 &  78.2 $\pm$ 2.1 & 8.64 &  0.0089 $\pm$ 0.0017 &      8.9 &     0.93 $\pm$ 0.01 & $<$ 1e-10 & Y \\
1046$-$409 &  130 &  1.5 &  6.0 &  23.4 $\pm$ 0.8 & 8.64 &  0.0021 $\pm$ 0.0012 &      2.1 &     0.95 $\pm$ 0.03 & 5.32e-03 & Y \\
1145$-$676 &   70 &  2.0 & 40.0 &  59.4 $\pm$ 1.8 & 8.64 &  0.0015 $\pm$ 0.0012 & $<$  5.1 & $>$ 0.95            & 3.74e-02 & N \\
1202$-$262 &  -15 &  1.5 &  8.0 &  60.3 $\pm$ 0.9 & 8.64 &  0.0225 $\pm$ 0.0022 &     22.5 &     0.86 $\pm$ 0.01 & $<$ 1e-10 & Y \\
1258$-$321 &  -70 &  1.5 & 20.0 & 122.5 $\pm$ 1.7 & 8.64 &  0.0026 $\pm$ 0.0012 &      2.6 &     1.03 $\pm$ 0.03 & 6.35e-04 & Y \\
1343$-$601 & -120 &  1.5 & 11.0 & 156.3 $\pm$ 2.5 & 8.64 &  0.0045 $\pm$ 0.0013 &      4.5 &     1.01 $\pm$ 0.02 & 2.46e-08 & Y \\
1424$-$418 & -110 &  1.0 &  4.0 &  40.6 $\pm$ 0.7 & 8.64 &  0.0018 $\pm$ 0.0015 & $<$  6.3 & $>$ 0.91            & 4.85e-02 & N \\
1655$+$077 &  -50 &  1.5 &  5.0 &  21.7 $\pm$ 0.6 & 4.86 & -0.0006 $\pm$ 0.0007 & $<$  1.4 & $>$ 0.93            & 9.18e-01 & N \\
1655$-$776 &   75 &  1.5 & 15.0 & 295.0 $\pm$ 3.6 & 8.64 &  0.0006 $\pm$ 0.0008 & $<$  3.0 & $>$ 1.07            & 1.53e-01 & N \\
1828$+$487 &  -40 &  1.0 &  2.5 & 272.8 $\pm$ 3.6 & 4.86 &  0.0278 $\pm$ 0.0030 &     27.8 &     0.91 $\pm$ 0.01 & $<$ 1e-10 & Y \\
2052$-$474 &  -25 &  1.5 &  4.0 &  10.3 $\pm$ 0.5 & 8.64 &  0.0004 $\pm$ 0.0007 & $<$  2.5 & $>$ 0.89            & 2.38e-01 & N \\
2101$-$490 &  100 &  1.5 & 16.0 &  74.9 $\pm$ 1.4 & 8.64 &  0.0031 $\pm$ 0.0009 &      3.1 &     0.99 $\pm$ 0.02 & $<$ 1e-10 & Y \\
2251$+$158 &  -50 &  1.5 &  7.0 & 278.0 $\pm$ 5.6 & 4.86 &  0.0142 $\pm$ 0.0029 &     14.2 &     0.95 $\pm$ 0.01 & $<$ 1e-10 & Y \\
\enddata
\tablenotetext{a}{The jet radio flux density is computed at $\nu_{\rm r}$ for the same region as for
the X-ray count rate, given by the PA, $r_i$, and $r_o$ parameters.  The X-ray flux
density is given at 1 keV assuming a conversion of 1 $\mu$Jy/(count/s), which is good
to $\sim$ 10\% for power law spectra with low column densities and spectral indices
($\alpha_{\rm x}$) near 1.5.}
\tablenotetext{b}{The quantity $P_{jet}$ is defined as the chance that there are more
counts than observed in the specified region under the null hypothesis that the counts
are background events.}
\tablenotetext{c}{The jet is defined to be detected if $P_{jet} < 0.0025$ (see text).}
\end{deluxetable}

\begin{deluxetable}{rrrrrrrr}
\tablecolumns{8}
\tablewidth{0pc}
\tabletypesize{\scriptsize}
\tablecaption{Quasar Jet Beaming Model Parameters \label{tab:beaming} }
\tablehead{
\colhead{Target} & \colhead{$\alpha_{rx}$} & \colhead{$R_1$\tablenotemark{a}}
	& \colhead{$V$\tablenotemark{b}} & \colhead{$B_1$\tablenotemark{c}} &
	\colhead{$K$\tablenotemark{d}} & \colhead{$\theta$\tablenotemark{e}} &
	\colhead{$\nu_{\rm s}$\tablenotemark{f}} \\
\colhead{ } & \colhead{}  & \colhead{($10^{-3}$)} & \colhead{(pc$^3$)} &
	\colhead{($\mu$G)} & \colhead{} & \colhead{(\arcdeg)} & \colhead{(MHz)} }
\startdata
0208$-$512 &     0.92 &      132.8 &  4.2e+15 &   7.4 &       2.34 &       20 & 10.0 \\
0229$+$131 & $>$ 0.95 &  $<$  55.8 &  7.8e+17 &   1.9 & $<$   0.15 &  $>$  40 & 26.1 \\
0413$-$210 &     1.04 &       13.0 &  7.0e+14 &  17.7 &       1.89 &       21 & 32.6 \\
0745$+$241 & $>$ 0.88 &  $<$ 230.3 &  2.9e+13 &  11.1 & $<$   9.88 &  $>$  13 &  5.0 \\
0858$-$771 & $>$ 0.99 &  $<$  40.3 &  6.4e+13 &  14.1 & $<$   4.25 &  $>$  17 & 14.1 \\
0903$-$573 &     1.07 &       10.1 &  3.6e+14 &  21.5 &       2.28 &       20 & 34.9 \\
0920$-$397 &     1.00 &       29.8 &  3.6e+14 &  13.9 &       3.09 &       18 & 17.9 \\
1030$-$357 &     0.93 &      103.0 &  1.9e+17 &   3.8 &       0.67 &       28 & 14.5 \\
1046$-$409 &     0.95 &       80.0 &  2.0e+14 &  10.4 &       3.84 &       17 & 10.5 \\
1145$-$676 & $>$ 0.95 &  $<$  77.7 &  3.6e+16 &   4.3 & $<$   1.01 &  $>$  25 & 13.5 \\
1202$-$262 &     0.86 &      335.5 &  1.3e+15 &   9.5 &       6.32 &       15 &  5.3 \\
1258$-$321 &     1.03 &       18.9 &  4.2e+08 &  75.6 &      33.47 &        9 & 14.0 \\
1343$-$601 &     1.01 &       25.8 &  9.1e+07 & 106.9 &      56.70 &        7 & 11.7 \\
1424$-$418 & $>$ 0.91 &  $<$ 140.6 &  5.4e+16 &   4.7 & $<$   0.92 &  $>$  25 & 12.6 \\
1655$+$077 & $>$ 0.93 &  $<$  94.7 &  1.6e+14 &   9.6 & $<$   3.87 &  $>$  17 &  9.6 \\
1655$-$776 & $>$ 1.07 &  $<$   9.1 &  9.5e+10 &  56.6 & $<$  14.25 &  $>$  12 & 22.9 \\
1828$+$487 &     0.91 &      145.3 &  1.3e+14 &  22.5 &      10.55 &       13 &  7.9 \\
2052$-$474 & $>$ 0.89 &  $<$ 214.4 &  3.8e+16 &   3.4 & $<$   0.88 &  $>$  26 &  9.8 \\
2101$-$490 &     0.99 &       37.6 &  1.4e+16 &   6.1 &       0.95 &       25 & 20.3 \\
2251$+$158 &     0.95 &       72.9 &  1.9e+15 &  12.3 &       3.23 &       18 & 12.9 \\
\enddata
\tablenotetext{a}{$R_1$ is the ratio of the inverse Compton to synchrotron
   luminosities, given by eq.~\ref{eq:ratio1}. }
\tablenotetext{b}{$V$ is the volume of the synchrotron emission region.}
\tablenotetext{c}{$B_1$ is the minimum energy magnetic field, given by
   eq.~\ref{eq:b1}.}
\tablenotetext{d}{$K$ is a function of observable and assumed quantities
   given by eq.~\ref{eq:kdef}.}
\tablenotetext{e}{The bulk Lorentz factor is assumed to  be 10.  This
	assumption would not be appropriate for 1343$-$601 because the extended
	material is not likely to be relativistic.}
\tablenotetext{f}{The frequency $nu_{\rm s}$ gives the part of the synchrotron
spectrum radiated by electons that upscatter CMB photons to 1 keV.}
\end{deluxetable}

\end{document}